\documentclass[
 amsmath,amssymb,
 aps,
]{revtex4-2}

\usepackage[utf8]{inputenc}
\usepackage{times}
\usepackage{natbib}
\usepackage{xcolor}

\usepackage{graphicx}
\usepackage{amsmath,amssymb}
\usepackage{gensymb,lipsum}
\usepackage{multirow}
\usepackage{verbatim}
\usepackage{dcolumn}
\usepackage{bm}

\begin{document}
\preprint{APS/123-QED}

\title{Helmholtz Thermodynamics Beyond Hamiltonians:\\ Including Walls, Pressure and Heat Flow }

\author{Amilcare Porporato}
 \affiliation{Department of Civil and Environmental Engineering and High Meadows Environmental Institute, Princeton University, Princeton, USA}


\author{Lamberto Rondoni}
\affiliation{%
Department of Mathematical Sciences, Politecnico di Torino, Corso Duca degli Abruzzi 24, 10129 Torino, Italy \\ INFN, Sezione di Torino, Via P. Giuria 1, 10125 Torino, Italy \\
ORCID: 0000-0002-4223-6279
}%

\date{\today}

\begin{abstract}
For 1D Hamiltonian systems with periodic solutions, Helmholtz formalism provides a tantalizing interpretation of classical thermodynamics, based on time integrals of purely mechanical quantities and without need of statistical description. Here we extend this approach to include heat flux and pressure at the walls, thereby enabling it to describe actual thermodynamic transformations, such as isothermal compressions and expansions. The presence of hard walls, which gives rise to non zero pressure, is justified by means of the virial theorem, while the heat fluxes are introduced as quasi-static limits of suitably thermostatted Hamiltonians. Particular attention is paid to generalizing the minimalist cases of the harmonic oscillator and elastic bouncer, which afford clear physical interpretations. With such extensions, a complete picture of thermodynamics emerges, amenable to 
cyclic transformations capable of producing mechanical work from heat, like the Carnot cycle.
\end{abstract}


\maketitle

\section{Introduction}

In 1884, Helmholtz showed that a thermodynamic formalism naturally emerges by simply taking temporal averages of solutions of one-dimensional mechanical systems over their oscillation periods, without need to resort to any probabilistic concept (cf.\ papers CXV through CXIX in Ref. \cite{He95}). After its extension by Boltzmann to $n$ dimensional systems (cf.\ the paper {\em ``\"Uber die Eigenschaften monozyklischer und anderer damit verwandter systeme''}), in Ref. \cite{Bo84}, this approach remained almost forgotten, until it resurfaced more recently, in the context of a search for dynamical foundations of statistical mechanics \cite{brush1976kind,gallavotti1999statistical,cardin2004helmholtz,campisi2005mechanical,campisi2010derivation}.

Helmholtz pioneering approach is still particularly relevant today, given the
continuing debate on conceptual issues involving the foundations of thermodynamics and the very definition of entropy in the phase space \cite{Andrey,RonEvan,ruelle2003extending,dunkel2014consistent,swendsen2016negative}
and, perhaps even more importantly, because present science and technology concern systems that exceed the boundaries of standard thermodynamics, and the search for suitable extensions is open. In particular, we have mechanical definitions for thermodynamic quantities such as pressure and temperature, but an equivalent accepted definition of entropy is still missing; entropy is in fact either derived from the former quantities by macroscopic considerations (e.g., Clausius), axiomatically (e.g., Caratheodory, Callen, etc.), or based on statistical concepts (e.g., Boltzmann and Gibbs). On the contrary, even if only for 1D Hamiltonian systems, in Helmholtz thermodynamics entropy is simply a time integral of the phase space area enclosed by the periodic trajectory.

In this paper, we revisit Helmholtz thermodynamics, extending it beyond the Hamiltonian formalism, to allow heat exchanges between system and environment. Even remaining within the confines of a 1D system, this line of research contains several elements that are useful to help strengthen and clarify the links between thermodynamics and mechanics. Many mechanical features of such systems are shared by any system of $N$ particles, and do not need to be supplemented with external hypotheses, whose applicability is often limited. With the goals of achieving a thermodynamic formalism that connects more directly to the thermodynamic of real systems, we explicitly add rigid walls that produce non zero pressure as well as include non conservative forces which give rise to energy exchanges in the form of heat. 

The first of these extensions becomes necessary because 1D potentials confine naturally the system in a way that no force is produced at the boundary, thus always resulting in zero pressure. As a consequence, while work can be done in theory by operating on a parameter of the potential, in practice the absence of a mechanical work term, which is formally similar to the $pdV$ mechanical work in simple thermodynamic systems, hinders a direct thermodynamic analysis featuring real expansions and compressions as those familiar in thermodynamics. 
On the other hand, hard walls bring about discontinuous velocities, produce non-zero forces at the boundary, and give rise to actual pressure, when such forces are averaged in time.

An even more important extension is related to the fact that Helmholtz thermodynamics is currently limited to Hamiltonian formulations, which do not allow for heat exchanges. Therefore, in the original Helmholtz thermodynamics quasistatic transformations remain adiabatic, 
with
with neither heat nor entropy exchanges. This considerably limits the scope of Helmholtz thermodynamics as a practical theory for real systems. To overcome this limitation, here we extend the theory to include non Hamiltonian components, opening the door to transformations with heat exchange and non-adiabatic (e.g., isothermal) transformations even in quasi-static conditions. The original Helmholtz approach requires the motion to be periodic. As we are interested in time averaged quantities, we will see how this limitation can be overcome, and non periodic systems considered. Going beyond adiabatic invariants (as in the theory by Einstein and Hertz) for slow transformations in Helmholtz thermodynamics, allows us to perform quasistatic changes in entropy with heat transfer and therefore to achieve a complete thermodynamic picture capable of defining material properties and transformation of heat into work by Carnot and similar thermodynamic cycles. 

The paper is organized as follows. After this introduction, Sec.\ 2 offers a brief review of Helmholtz thermodynamics and its connection  with classical mechanics. Sec.\ 3 further exploits this link to formulate a thermodynamic equations of state. Sec.\ 4 goes into the details of the limitations of a Hamiltonian picture without rigid walls. Sec.\ 5 uses the virial theorem to define pressure produced by rigid walls, while Sec.\ 6 illustrates the introduced extensions via examples. Sec.\ 7 adds the interactions with the environment and therefore the heat fluxes. Sec.\ 8 presents a Carnot cycle performed with the proposed extension of the Helmholtz theory. Finally, Sec.\ 9 draws some conclusions and discusses open problems.

\section{From Hamiltonian Mechanics to Helmholtz  Thermodynamics}

Helmholtz considered one-dimensional Hamiltonian systems
\begin{eqnarray}
    &&\dot{x}=\frac{p}{m} \nonumber \\
    && \label{dynsys} \\
    &&\dot{p}=-\partial_x \Phi(x,\lambda(t)), \nonumber
\end{eqnarray}
where the conservative force is linked to a potential $\Phi(x,\lambda(t))$, which may either be thought of as an external field acting on the system or as the inter-particle interaction energy (see e.g., \S{13} of \citep{landau1976mechanics}).  Accordingly, $x$ is either the position of the single particle or the distance between interacting particles. The potential may depend on a set of parameters, $\lambda$, possibly dependent on time. Convex potentials give rise to systems, called monocyclic or hortodes, with periodic solutions. For such systems Helmholtz  proposed a thermodynamic interpretation where the internal energy is linked to the Hamiltonian,
$$
\mathcal{H}=K+\Phi=E,
$$
where $K$ is the kinetic energy and $E$ is the total energy.

Entropy is assumed proportional to the logarithm of the 2D phase space area enclosed by the periodic orbit. The latter is 
\begin{equation}
\Omega = m \oint | v | ~  {\rm d} x 
\quad \mbox{where } ~~
 |v(x ; \lambda)|=\sqrt{\frac{2}{m} (E-\Phi(x ; \lambda))}
 \label{absv}
\end{equation}

which can be written as:
\begin{equation}
\Omega(E ; \lambda)=2 \int_{x^-}^{x^+} \sqrt{2m (E-\Phi(x;\lambda))} \, dx 
\label{eq:Omega}=\int_0^{\tau(E ; \lambda)} m v^2 dt
\end{equation}
where $x^{-}$ and $x^{+}$ are the extreme positions reached during the oscillation and $\tau(E;
\lambda)$ is the period,

which is linked to the absolute value of the 
velocity as:
\begin{equation}
    \tau(E ; \lambda)=\sqrt{2m} \int_{x_-}^{x_+} \frac{dx}{\sqrt{E-\Phi(x ; \lambda)}} =  \frac{\partial \Omega}{\partial E}
    \label{eq:period}
\end{equation}
Following Helmholtz, temperature may be formally introduced  {\em via} the
mean kinetic energy over a period; the phase space area is also the time integral of the kinetic energy over the period, one obtains:
\begin{equation}
k_B T=\frac{2}{\tau(E ; \lambda)}
\int_0^{\tau(E ; \lambda)} \frac{1}{2}m v^2 dt
= \frac{1}{\tau(E ; \lambda)}\int_0^{\tau(E ; \lambda)} m |v| 
\left|\frac {d x}{d t}\right| dt
=\frac{1}{\tau(E ; \lambda)} \Omega(E ; \lambda)
\label{eq:temp}
\end{equation}
Then, a formal definition of entropy is expressed by:
\begin{equation}
S(E;\lambda)=k_B \ln 2 \int_{x^-}^{x_+} \sqrt{2m (E-\Phi(x;\lambda))} dx = k_B \ln \oint |p| dx =k_B \ln \Omega(E;\lambda)
\label{ENTR}
\end{equation}

Hence the thermodynamic definition of temperature follows, if $E$ is interpreted as internal energy, and $S$ as entropy:
\begin{equation}
    \frac{\partial E}{\partial S}=T.
    \label{eqTl}
\end{equation}
The ``pressure'' may now be defined as the time average of 
$-{\partial \Phi}/{\partial \lambda}$, 
\begin{equation}
P_\lambda = -\frac{1}{\tau}\int_0^\tau \frac{\partial \Phi}{\partial \lambda} dt.
\label{eq:pressdef}
\end{equation}
The quotes are meant to emphasize that, unless $\lambda$ is a distance, the integrand is not a force (which must then be referred to a unit surface, to yield a pressure). Hence, more properly, $P_\lambda$ is to be interpreted as the intensive quantity conjugated to the parameter $\lambda$, whatever physical quantity that may be, {\em e.g.}\ the elastic constant of a harmonic potential.

With these definitions, the Helmholtz theorem finally shows that the above mechanical quantities are consistent with the thermodynamic formalism, in the sense that they obey analogous relations, such as Eq. \eqref{eqTl}, and 
\begin{equation}
\frac{\partial E}{\partial \lambda}=\frac{\partial \Phi}{\partial \lambda}=-P_\lambda,
\end{equation} 
which eventually lead to the Gibbs equation:
\begin{equation}
dE=TdS-P_\lambda d\lambda. 
\end{equation}

\section{From Classical Mechanics to the Thermal Equation of State}

For systems in which the energy $E$ is constant (and therefore the parameter
$\lambda $ is also a constant), the Hamilton-Jacobi equation for the action $A(t)$ can be solved to get \citep[p.~444]{goldstein2002classical}
\begin{equation}
A(t ; \lambda)=\int_{0}^{t}\mathcal{L}(t' ; \lambda)dt'=-E t+W(t ; \lambda),
\label{eq:action}\end{equation}
where 
\begin{eqnarray}
\mathcal{L}=K-\Phi
\end{eqnarray}
is the Lagrangian and 
$$
W(t)=m \int_0^t v^2 dt' \, , \quad \mbox{with} ~~~
 \ m\int_0^t v^2 dt' = 
\int_{x_0}^{x(t)} |p| dx' \, ,
$$ 
is the abbreviated action, which corresponds to the area enclosed by the trajectory in the plane 
$\{x,p\}$ up ot time $t$.

For a period $\tau$, the abbreviated action is the entire area enclosed by the closed orbit, $\Omega=2\pi I$, where $I$ is the action variable. As a result,  the previous equation may also be written as the time average of the Lagrangian:
\begin{equation}
-\overline{\mathcal{L}}(E ; \lambda)= -{A(\lambda) \over \tau} = E- {\Omega(\lambda) \over \tau}=E-2 \overline{K}.
\end{equation}
Using the kinetic definition of temperature (\ref{eq:temp}),  
one obtains
\begin{equation}
E=k_B T-\overline{\mathcal{L}}(E;\lambda).
\end{equation}
Thus this equation plays the role of a thermal equation of state (see Callen \cite{callen1998thermodynamics}, p. 37 and 63).

\section{Need for Extending Helmholtz Thermodynamics}

While Helmholtz thermodynamics offers a very powerful link to thermodynamics, as mentioned in the introduction, there are two issues that limit its direct contact with thermodynamics, the lack of a proper pressure and the absence of heat fluxes. These are discussed in more detail next.

\subsection{Adding Walls to Define Volume and Proper Pressure}

A first general limitation of Helmholtz thermodynamics is that, in general, $P_\lambda$ does not have the units of a pressure, since $\lambda$ can be any parameter of the potential, and thus the term $ P_\lambda d \lambda $ does not in general behave as the mechanical reversible work $PdV$ in the Gibbs equation for simple thermodynamic systems. Therefore, $P_\lambda$ should be considered just as a generalized intensive quantity conjugated to $\lambda$. 
To go beyond this, pressure should be conjugated to a quantity that would serve as a volume (i.e., a length in 1D). As an initial candidate for this length, one could consider the amplitude of the oscillations, $\ell=x^{+}-x^{-}$, that is determined by the energy $E$ of the system and by the value of the parameter $\lambda$. This can
be obtained from the period of the oscillation, which in turn is linked to the absolute value of the 
velocity, cf.\ \citep[Eq.~(12.1)]{landau1976mechanics}, and Eq.\ \eqref{absv} and \eqref{eq:period}:

\begin{equation}
    \ell (E ; \lambda)=x_+-x_-=\frac{1}{\pi\sqrt{2m}}\int_0^E \frac{\tau(E' ; \lambda)}{\sqrt{E-E'}}dE'.
    \label{elle}
\end{equation}
The system is indeed bound to stay within the interval $[x^-,x^+]$.
However, taking $\ell$ as the volume always yields a vanishing pressure, because, without further specification, the ends of the interval are the points at which the kinetic energy vanishes, but are not barriers beyond which it is impossible to go. As a result, adopting $\ell$ as the volume would not provide the needed independent variable; in fact, were an object placed at $x^-$ or $x^+$, {\em e.g.}\ a piston, it would feel no force exerted by the system. That this is so, in a thermodynamic interpretation, is also clear from the fact that thermodynamic pressure $P$, internal energy $U$ and volume $V$ obey:
\begin{equation}
-P = \cfrac{\partial U}{\partial V}
\end{equation}
but interpreting $E$ as the internal energy and $\ell$ as the volume, we would
have:
\begin{equation}
-P = \left( \cfrac{\partial \ell}{\partial E}\right)^{-1} =
\left. \cfrac{\pi \sqrt{2m(E-E')}}{\tau(E' ; \lambda)} \,\, \right|_{E'=E} = 0
\end{equation}

To proceed further, it is important to distinguish three kinds of divergences of potentials, which refer to different physical situations. The first represents an external field, that is not affected by the motion of the particles, and grows as the distance does, diverging in the limit of large distances. In this case, an increase of the system energy leads to a larger domain (like $\ell$, in the case above) that particles can explore; that is, a volume as large as necessary for the pressure exerted on a hypothetical object placed at the ends of such volume to vanish.
This implies that, unlike systems bound by walls, particles with different energy within the same system, or a given particle whose energy changes, are confined in regions of different volume (exploring the role of rigid walls in the pressure definition would also be interesting the in the thermodynamics interpretation of stochastic models of active particles \cite{solon2015pressure}). 
The second represents repulsive interactions among particles, like hard core or Lennard-Jones potentials; the divergence occurs at small distances, but its location changes in time as it depends on the motion of the particles, and it does not delimit a specific region of space. The third more properly represents walls placed at given positions in space, {\em e.g.}\ at a fixed distance $L$ from each other, not dependent on the energy. Also, the potential does not affect the motion of particles unless they are very close to the divergence points, and the potential itself is not affected by the particles. 
In this case, particles bouncing back after colliding with a wall exert a net force at the boundary of the volume $L$: the variation of momentum at collisions  determines the average force, hence the pressure.
Since case 2 of finite distance divergence in interparticle collisions does not generate pressure, {\em per se}, as equilibrium may not be reached in absence of a confining finite volume, this third case of a potential representing a wall must thus be distinguished from both external fields and particles interactions.
Of course, hard walls could be replaced by --perhaps more realistic--
very steep potentials, that do not distinguish {\em in practice} particles of different energy. This, however, merely complicates the formalism with no gain for the development of the theory.  For instance, one would have to exclude particles with energy higher than a given threshold, and specify which penetrations of the potential can be neglected, so that the results implied by hard walls be negligibly perturbed.

\subsection{Allowing for Heat Flux}

A second major limitation of the traditional approach based only on Hamiltonians is due to the fact that this representation, by construction, does not include heat transfer. In this case, in fact, $\Omega$ is an adiabatic invariant \citep{landau1976mechanics}, which is to say that for time dependent parameter $\lambda=\lambda(t)$, it remains constant if the transformation is very slow. 
Since $S$ does not change either, one obtains
\begin{equation}
dE=-P_\lambda d\lambda,
\end{equation}
which means that reversible work can be done in quasi-static conditions but that there is no reversible heat transfer.
Thus, to comply with thermodynamics and to include heat transfer and allow for more general transformations ({\em e.g.}\ Carnot cycles), Helmholtz theory needs to be extended. 

In the context of linking mechanics and thermodynamics, it is important to remark that the term adiabatic has the potential to create some ambiguity if not used carefully. In particular, it is important to distinguish the meaning of adiabatic invariance in classical mechanics from the one of adiabatic transformation in thermodynamics. In fact, the term adiabatic in thermodynamics means `occurring without loss or gain of heat' from the Greek term `impassable' and refers to transformations that are not necessarily quasi-static, but that may be done in finite time. Differently, in classical mechanics adiabatic invariant is used to refer to quasi-static transformations of Hamiltonian systems (quasi-static + no heat flux), and therefore refer only to a subset of the adiabatic transformations considered in thermodynamics (no heat flux). This distinction should be borne in mind to avoid confusion.

\section{Virial Theorem, Pressure and the Mechanical Equation of State for  Isolated System}
As well known, the virial theorem is a purely mechanical, generally valid result, concerning the time averages of the kinetic energy and of the product of force and position (not force times displacement, hence not mechanical work). For our goal here, the virial theorem provides a way to introduce a proper pressure and derive a mechanical equation of state, which directly links it to thermodynamics.
In order to introduce a proper pressure that, unlike $P_\lambda$ 
defined by Eq. (\ref{eq:pressdef})), does not vanish at equilibrium, we first need to introduce rigid walls at a distance $L$. This quantity will serve as the 1D volume confining our system, providing a way to act on the system by changing its size $L$, independently of the energy, as done by moving the position of a piston in a $pVT$ thermodynamic system.

Given the initial condition $(x_0,p_0)$ of a trajectory in the phase space of a system with one degree of freedom, and denoting by a bar the infinitely long time average (or similarly the average over a period, when appropriate), the theorem states:
\begin{equation}
\overline{K\,} (x_0,p_0;L,\lambda) = 
- \frac{1}{2} \,  \left[ \, \overline{ x \, F^{\rm wall}_L} \, (x_0,p_0) +
\overline{ x \, F^{\rm int}_\lambda} \, (x_0,p_0)  
\right]
\label{VirThmDec}
\end{equation}
where the force has been split in one part due to a wall, $F^{\rm wall}_L$, which diverges at two points separated by distance $L$, say $x=-L/2$ and $x=L/2$, and the other part due to interactions of other sort, depending on a set of parameters $\lambda$, $F^{\rm int}_\lambda$. Note that in principle the dependence on the initial condition remains even if the averaging time is infinite. However, considering 1-dimensional periodic systems, as we do here, that dependence disappears if one averages over a single period or multiples of that.

Assuming that the particle collides with walls of unit surface, at the two ends of the segment of length $L$, a uniform pressure $P$ arises as 
\begin{equation}
\overline{K\,} (L,\lambda) = 
- \frac{1}{2} \,  \left[ \, \overline{ x \, F^{\rm wall}_L} \, (x_0,p_0) +
\overline{ x \, F^{\rm int}_\lambda} \, (x_0,p_0)  
\right] = \frac{1}{2}  \left[-\frac{L}{2} \cdot  P + \frac{L}{2} \cdot P -  \overline{ x \, F^{\rm int}_\lambda} \, \right]
= \frac{1}{2} \left[ \, P L  -  \overline{ x \, F^{\rm int}_\lambda} \,
\right]
\label{VirThmDecLl}
\end{equation}
where, 
$P$ is the time averaged force per unit area localized at the walls, and $L$ represents the volume. With the average kinetic energy being related to the temperature $T$, as in the equilibrium cases in which the energy equipartition holds, we can write:
\begin{equation}
2 \overline{K\,} + \overline{ x \, F^{\rm int}_\lambda} 
= 2 \overline{K\,} + \mathcal{V}
=  P L \, , \quad \mbox{or } \,\,\,
P = \frac{1}{L} k_{_B} T + \frac{1}{L} {\mathcal{V}},
\label{VirThmDekb}
\end{equation}
which defines the interaction term $\cal V$ and constitutes a mechanical equation of state, analogous to the thermodynamic one.
In the case there is no wall, all forces are represented by $F^{\rm int}_\lambda$, $\cal V$ is the virial of Clausius, and the pressure is computed to vanish also along this path, because 
\begin{equation}
    \overline{K} = - {1 \over 2} \, \cal V.
\end{equation}
If the forces are expressed via their potentials:
$$
F^{\rm wall}_L = - \frac{\partial}{\partial x} \Phi^{\rm wall}_L \, ; \quad \mbox{and } ~~~
F^{\rm int}_L = - \frac{\partial}{\partial x} \Phi^{\rm int}_\lambda,
$$
we can write
\begin{equation}
PL = 2 \overline{K} - \overline{ \, x \,
\frac{\partial \Phi^{\rm int}_\lambda}{\partial x} \,},
\label{playgc}
\end{equation}
which plays the role of the grandcanonical or Landau potential, $PV$, where $V=L\cdot 1$. 
When $\Phi^{\rm int}_\lambda=0$, or its time average is negligible, since $x$ is bounded, we recover the ideal gas law; otherwise Eq.\ \eqref{playgc} corresponds to real gases. One should observe that, for the time average to be negligible, the quantity  $\Phi^{\rm int}_\lambda$ may also have divergences, as long as they last only a short fraction of time. That is what allows Boyle's law to hold, despite the fact that, even in a perfect gas, molecules have to interact for thermodynamic fields to make sense
\cite{landau2013course,rondoni2002some,rondoni2021introduction}. 

In case all forces derive from time independent potentials, the system  exchanges no energy with the outer environment, and one may speak of an 
adiabatic system. In such a case, the energy $E$ is constant, while the kinetic energy, like other quantities, fluctuates and its
time average is determined by the chosen total energy: 
$$\overline{K}=\overline{K}(E,L,\lambda)  
\quad 
\mbox{and } \quad
\overline{\Phi}=\overline{\Phi}(E,L,\lambda).
$$ 
With the positive pressure, $P$, determined by walls at a distance $L<\ell$, the same definitions as in Helmholtz thermodynamics may be adopted. As a result, the formalism persists qualitatively untouched and with the pressure so defined the system can now truly do work as force times displacement, as in the thermodynamics of PVT systems by moving a piston. Quasistatic transformations are isentropic, and the Gibbs equation becomes
$$
dE=-PdL,
$$
assuming $\lambda$ remains constant, while entropy is still an adiabatic invariant for very slow changes in $L$.

\section{Special Cases}

The detailed discussion of some simple special cases that often can be solved analytically is useful to illustrate the theory and its extension. We emphasize in particular the precise thermodynamic formalism that follows from Helmholtz thermodynamics, from the fundamental equations all the way to the material properties.

\subsection{Elastic Bouncer}

The elastic bouncer \cite{arnol2013mathematical} is perhaps the simplest case which affords a complete thermodynamics, paralleling the one of an ideal gas, if hard walls are introduced. 
In this case, $\Phi^{\rm int}=0$, hence the speed  $v$ of the particle is constant, and the period of the motion is $\tau = 2L/v$, if the walls are placed at a distance $L$. Clearly, an equilibrium exists only if walls at a finite distance are introduced. Using the equations previously derived, one has:
\begin{equation}
   P = \cfrac{1}{L} ~ k_{_B} T \, ; \quad E(T) = \cfrac{k_{_B}T}{2} \, ; \quad C_v = \cfrac{k_{_B}}{2}.
\end{equation}
These equations correspond to the ones of an ideal monoatomic gas in one dimension, {\em if} walls define a finite volume within which a periodic motion takes place.

Recalling that the action variable is $I=\Omega/(2\pi)=\sqrt{2m \mathcal{H}}L/\pi$, so that $\omega=\partial \mathcal{H}/\partial I=\pi^2I/(m L^2)=\pi v/L$, one obtains:
\begin{equation}
    \Omega= 2\pi I = \sqrt{8m \mathcal{H}}L = 2mv L = 2 \sqrt{2 m E} ~ L = 2 L \sqrt{m k_{_B}T}
\label{BouHJ}
\end{equation}
As a result, the entropy becomes:
\begin{equation}
    S(T,L) = \cfrac{k_{_B}}{2} ~  \ln T + k_{_B} \ln L + \mbox{const}
\end{equation}
or, in the form of a fundamental equation as a function of energy and volume \cite{callen1998thermodynamics}, 
\begin{equation}
    S(E,L) = \cfrac{k_{_B}}{2} ~  \ln E + k_{_B} \ln L + \mbox{const}
\end{equation}
which yields
\begin{equation}
   \left.  \cfrac{\partial S}{\partial E} 
   \right|_L = \cfrac{1}{T} \, ; \quad 
    \left.  \cfrac{\partial S}{\partial L} 
   \right|_E = \cfrac{k_{_B}}{L}=\cfrac{P}{T} \, ,
\end{equation}
as one would expect from the thermodynamic formalism \cite{callen1998thermodynamics}. Then, defining the pressure as the mean force of the bouncer, when it hits the walls, we have 
\begin{equation}
    P=\frac{2vm}{\tau}=\frac{v^2 m}{L}=\frac{k_{_B}T}{ L} 
    ~ , 
\label{eq:pressbouncer}
\end{equation}
which directly yields the equation of state
$$
PL =  k_{_B} T.
$$
Finally, one can calculate the material properties \cite{callen1998thermodynamics}, including the coefficient of thermal expansion
\begin{equation}
    \alpha=\frac{1}{L}\left.\frac{\partial L}{\partial T}\right|_P=\frac{1}{T}
\end{equation}
and the isothermal compressibility
\begin{equation}
    \kappa_T=-\frac{1}{L}\left.\frac{\partial L}{\partial P}\right|_T=\frac{1}{P}
\end{equation}
which coincide with the ones of an ideal gas (see e.g.,  \cite{callen1998thermodynamics}, p.87).

\subsection{Harmonic Oscillator (No Walls)} 

We consider first a harmonic oscillator with no walls.
The spring constant, $k$, may be taken as the parameter of the potential, and the Hamiltonian writes:
\begin{equation}
    \mathcal{H} = \cfrac{m}{2} \, \dot{x}^2 + \cfrac{k}{2} \, x^2
\end{equation}
The corresponding frequency is $\omega=\sqrt{k/m}$, and the period of the motion the motion is $\tau=2 \pi/\omega$.
From the solution of the equations of motion, the amplitude of the oscillations is expressed by: 
\begin{equation}
\ell/2=\sqrt{x_0^2+\frac{v_0^2}{\omega^2}} =\sqrt{\frac{2E}{k}}
\end{equation}
while the phase $\varphi$ is not important, because thermodynamics arise only from averaging over the whole period.

To calculate the entropy, we first need
the area $\Omega$ delimited by the ellipse drawn by trajectories in phase space, which has semi-axes $s_x=\sqrt{2E/k}$ and $s_v=\sqrt{2mE}$, the area is
\begin{equation}
    \Omega=\pi s_x s_v=\frac{2\pi E}{\omega}=
\tau E 
\end{equation}
and therefore for the action variable 
\begin{equation}
    I=\frac{E}{\omega} 
\label{eq:actionharm}\end{equation}
Equation (\ref{eq:actionharm}) shows that 
\begin{equation}
    A_0=-E\tau+\Omega=-\frac{2\pi E}{\omega}+\frac{2\pi E}{\omega}=0
\end{equation}
This is related to the fact that the virial theorem in this simple case states that time averaged kinetic and potential energies are the same,
so that $E=k_{_B}T$. As a result,
\begin{equation}
    S(T) = k_{_B} \ln \Omega = 
    k_{_B} \ln T + k_{_B} \ln \tau +
    k_{_B} \ln k_{_B}
    = 
    k_{_B} \ln T + \mbox{constant}
\end{equation}
The pressure  is zero, since no walls have been introduced. In fact, all points of the phase-space ellipse have same energy $E$ and:
\begin{equation}
E = \overline{K}^0 + \overline{\Phi}_{k}^0 = 2 \overline{K}^0 \, , \quad \mbox{so } \quad \overline{K}^0(E) = 
  \cfrac{E}{2}  = \overline{\Phi}_{k}^0 \quad
  \mbox{and } \quad \overline{ \, x \,
\frac{\partial \Phi_k}{\partial x} \,}^0 = 2 \overline{K}^0(E) \, , \quad
\mbox{which implies} \quad P = 0
\end{equation}
Here, the apex 0 highlights the fact that there are no walls, {\em i.e.}\ $L=\ell$.
The value of $\ell$ is obtained imposing that the instantaneous kinetic energy vanishes, which yields: $\ell(E,k)=2\sqrt{2E/k}$. 

The fundamental relation in terms of temperature is:
\begin{equation}
   E(T) = 
    k_{_B} T,
\end{equation}
which has no dependence on the volume and on $k$. Thus, correctly, differentiation with respect to volume yields vanishing pressure. Differentiation with respect to temperature yields a heat capacity which is the same at constant (vanishing) pressure, because the pressure remains 0 at all temperatures,
\begin{equation}
 \cfrac{\partial E}{\partial T}  =  k_{_B}.
\end{equation}

\subsection{General Power-Law Potential}
Before adding rigid walls, it is interesting to note the effect of changing the potential on the thermal properties of the system, namely the heat capacity. For a general oscillator with potential energy of the form $\Phi(x)=A|x|^n$ the period is \cite{landau1976mechanics}
\begin{equation}
    \tau=\frac{2}{n}\sqrt{\frac{2\pi m}{E}}\left(\frac{E}{A}\right)^{1/n}\frac{\Gamma(1/n)}{\Gamma(1/2+1/n)}
\end{equation}
where $\Gamma$ is the gamma function.
Since ${\partial \Omega}/{\partial E}=\tau$, the previous result can be used to obtain the $\Omega(E)$ relationship, by solving
\begin{equation}
    \frac{\partial \Omega}{\partial E}=\frac{2}{n}\sqrt{\frac{2\pi m}{E}}\left(\frac{E}{A}\right)^{1/n}\frac{\Gamma(1/n)}{\Gamma(1/2+1/n)},
\end{equation}
which gives
\begin{equation}
    \Omega=2 \sqrt{2\pi m} A^{\frac{1}{n}}\frac{\Gamma \left(\frac{3}{2}+\frac{1}{n}\right)}{\Gamma \left(\frac{1}{2}+\frac{1}{n}\right)}E^{\frac{1}{2}+\frac{1}{n}}.
\end{equation}
As a result, the entropy is
$$
S=k_B \left(\frac{1}{2}+\frac{1}{n}\right) \ln E + \mbox{const}
$$
and the heat capacity
$$
C = k_B \left(\frac{1}{2}+\frac{1}{n}\right).$$
Without walls, also this case corresponds to a zero pressure state.

\subsection{Harmonic Oscillator with Walls} 

Introducing hard walls at $x^- = -L/2$ and $x^+ = L/2$, with $L < \ell(E,k)$, the period of the cycle for an harmonic oscillator is given by 

\begin{equation}
\tau(E;L,k) = \sqrt{2m} \int_{-L/2}^{L/2} \frac{dx}{\sqrt{E-\Phi(x ; \lambda)}} 
= \cfrac{4}{\omega_0} \arcsin\left(\cfrac{L}{2} \,{\sqrt{\cfrac{k}{2E}}} \, \right)=\tau^0\cfrac{2}{\pi} \arcsin\left(\cfrac{L}{2} \,{\sqrt{\cfrac{k}{2E}}} \, \right)
    \label{tauOHL}
\end{equation}
which reduces to $2\pi/\omega_0$ for $L=\ell$, as it should. In turn,
the average kinetic energy is now expressed by
\begin{equation}
\overline{K} = \cfrac{E}{2} ( 1 + Q_k ) =
\overline{K}^0 \left( 1 + Q_k \right) = 
E - \overline{\Phi}_k  > \overline{K}^0 
\label{KLmed}
\end{equation}
where:
\begin{equation}
Q_k = 
\cfrac{\sin \left( 2 \arcsin\left(\cfrac{L}{2} \,{\sqrt{\cfrac{k}{2E}}} \, \right) \right)}{2 \arcsin\left(\cfrac{L}{2} \,{\sqrt{\cfrac{k}{2E}}} \, \right)} =
1 - \cfrac{1}{2\overline{K}^0} ~ \,
\overline{x\cfrac{\partial \Phi_k}{\partial x}}
\, \quad 
\mbox{and } ~~
\overline{\Phi}_k = \cfrac{1}{2} ~
\overline{x\cfrac{\partial \Phi_k}{\partial x}}
< \overline{\Phi}_k^0
\label{KLmed1}
\end{equation}
because for $0 < L < \ell$, the arcsin lies in $(0,\pi/2)$ and $Q_k \in (0,1)$. Indeed, while the kinetic energy is averaged around its largest values, the potential energy is not, because positions do not reach their largest extension. Then, because the average of $\Phi_k$ is smaller than twice the average of $K$, Eq.\eqref{playgc} yields $P > 0$.
Interestingly, the maximum value of $\overline{K}$, which is $E$, is approached in the $L \to 0$ limit, under which 
$\overline {x \partial_x \Phi_k} \to 0$ and 
$P \to \infty$. Equation \eqref{KLmed}, with the definition $2\overline{K}=k_{_B} T$ yields:
\begin{equation}
    P = \cfrac{1}{L} \,  k_{_B} T^0 (1 + Q_k ) -  \cfrac{1}{L} \,  \overline{\, x\cfrac{\partial \Phi_k}{\partial x} \,}  =  k_{_B} T -  
    \cfrac{1}{L}\,  \mathcal{V}
\end{equation}
where $T> T^0$.

\begin{figure}
\includegraphics[width=208pt]{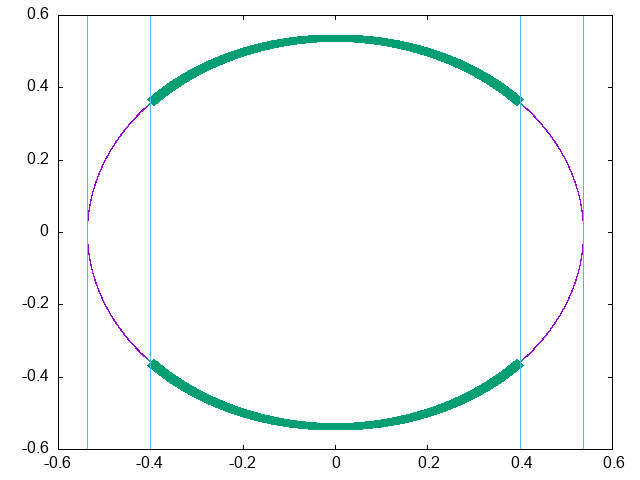} 
~~~
\includegraphics[width=208pt] {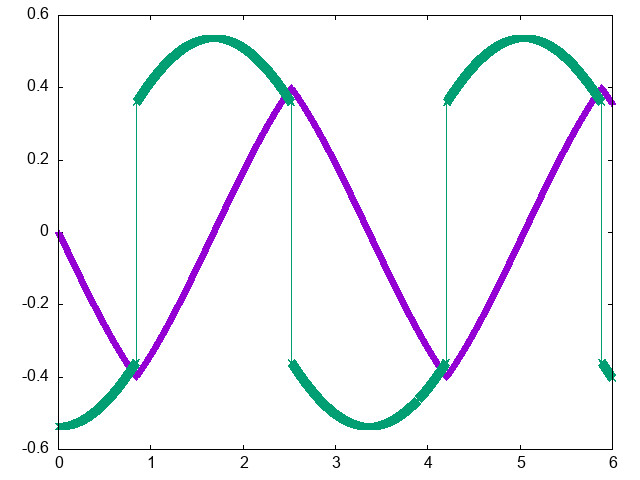}
\caption{Left panel: phase space trajectory of harmonic oscillator with walls at $x_\pm=\pm 0.4$, indicated by the two inner vertical lines (green points), and
phase space trajectory in absence of walls, limited by $\ell=0.54$, indicated by the outer vertical lines (purple line). Right panel:}
positions (purple points) and momenta (green points) as functions of time. Positions are continuous, with a cusp at the walls, where velocity is discontinuously inverted.
 \label{fig:wall}
\end{figure}

The solution of the equation of motion between two bounces with the walls (see Figure 1) is expressed by:
\begin{equation}
    x(t) = \sqrt{\cfrac{2E}{k}} ~ \cos({\omega t + \varphi}) ~ ; \quad
    \dot{x}(t) = 
    - \sqrt{\cfrac{2E}{m}} ~ \sin({\omega t + \varphi})
\end{equation}
where, having fixed the energy, we can take
time 0 corresponding to the maximum of the kinetic energy, which means 
$v(0) = \sqrt{2E/m}$ and $x(0)=0$, {\em i.e}\ $\varphi = \pi/2$. Then, the phase space area is given by:
\begin{equation}
    \Omega = 2 \int_{-\tau/2}^{\tau/2}
    m \dot{x}^2 d t = 2 m \int_{-\tau/2}^{\tau/2} \cfrac{2 E}{m} 
    \cos^2 ( \omega t ) ~  d t = 
    2 \tau \overline{K}=\tau T
\end{equation}
where $\overline{K}$ is given by Eq.\ \eqref{KLmed}. In turn,
\begin{equation}
    S = k_{B} \ln T^0 + k_{B} \ln Q_k(T,L,k) + k_{B} 
 \ln \tau(T,L,k) + \mbox{const}
\end{equation}
Thus, both $\Omega$ and the entropy acquire a dependence on $L$. 

The fundamental relation for the internal energy as a function of temperature $E=E (T,L,k)$ can be derived inverting Eq.\ \eqref{KLmed} with respect to $E$. Then, for the heat capacity one may proceed implicitly, differentiating $T$ with respect to $E$:
\begin{equation}
\cfrac{1}{C_v} = 
\left. \cfrac{\partial T}{\partial E} \right|_{L,k} = \cfrac{1}{k_{_B}} \left[
1 + Q_k(E,L,k) + E \cfrac{\partial Q_k}{\partial E}
\right],
\end{equation}
where the effect of the walls on the heat capacity becomes evident.

\subsection{Harmonic Oscillator Compressed by an Adiabatic Piston}   
We consider here the case in which the harmonic oscillator undergoes an adiabatic compression, by moving the right wall as if it
were an adiabatic piston. With no other perturbation involved, 
the energy of the harmonic oscillator, hence the size of its phase-space ellipse, increases under compression, because collisions with the moving wall enhance the particle's momentum by a fixed amount.
As the right wall nears the left wall,
the collisions get more frequent and the energy increases at a faster rate.
As seen in Figure 2, the behavior of the oscillations for decreasing $L$ tends to those of the elastic bouncer (with ever greater energy), as they involve relatively flat portions of the 
potential, 
and thus reproduces the ideal gas law.
However, the heat capacity decreases with temperature as the volume is decreased at constant energy, with an opposite behavior to the one of a real gas, where instead $C_v$ increases as  
temperature is increased. In the next section, we will see that this is due to the fact that the potential is attractive.  

\begin{figure}
\begin{center}
\includegraphics[width=208pt]{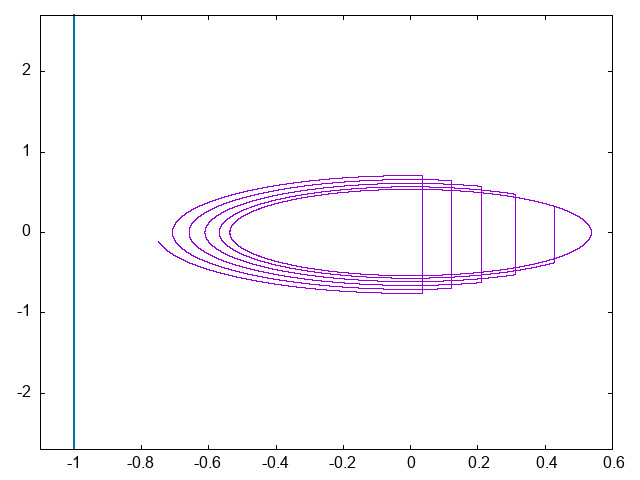} ~~
\includegraphics[width=208pt]{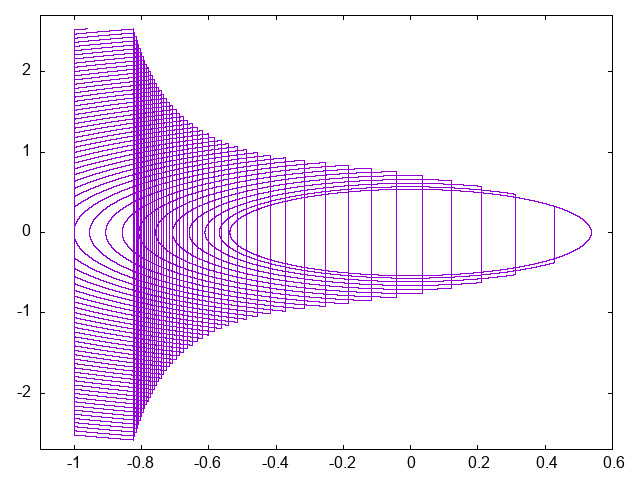}
\end{center}
\caption{Compressed adiabatic harmonic oscillator.
Left: initially the oscillator gains energy from the compressing piston (right wall slowly moving to the left), thus its ellipse grows in size remaining centered at the origin of axes, but being cut on the right. Right: when the ellipse hits the left wall, the energy increases more rapidly, as illustrated by the diverging momentum}.
\end{figure}

\subsection{Stepwise Repulsive Potential as a Minimalist Model of a Real Gas}

As a minimalist model of a real gas we consider a stepwise repulsive potential 
\begin{equation}
    \mathcal{H}=\left\{\begin{array}{lll}
        K+E^\alpha & {\rm if} & |x|<1/2 \\
        K & {\rm if} & |x| \in [1/2,L/2]
    \end{array}\right.
    \label{eq:steprealgas}
\end{equation}
for $\alpha<1$, whereby an elastic bouncer receives two kicks at $x=\pm 1/2$, directed toward the nearest wall. 
As a result of these kicks, the particle slows down as it enters the constant potential region $|x|<1/2$, where the velocity is $v_0=\sqrt{2(E-E^\alpha)/m}$, and accelerates as it enters the region $|x|\in [1/2,L/2]$, where it travels with speed $v_1=\sqrt{2E/m}$. 
The half period of the oscillation is  the sum of the times spent in each region, 
\begin{equation}
    \frac{\tau}{2}=t_0+t_1=\sqrt{\frac{m}{2E}}(L-1)+\sqrt{\frac{M}{2(E-E^\alpha)}},
\end{equation}
so that the average kinetic energy is
\begin{equation}
    \overline{K}=E-\left(\frac{\sqrt{E(E-E^\alpha}}{(E-E^\alpha)(L-1)+\sqrt{E(E-E^\alpha)}}\right).
\end{equation}
The resulting temperature and heat capacities are plotted in Figure 3 for different cases. In particular, the heat capacity increases monotonically with temperature, becoming constant for large temperatures, when the collisions with the step potential negligibly contribute to the particle's energy, and the ideal gas condition is recovered.
For $\alpha=0$, as well as in general for large $L$, the heat capacity  tends to the one of the ideal gas, ${2}/{k_B}$.
\begin{figure}
\begin{center}
\includegraphics[width=240pt]{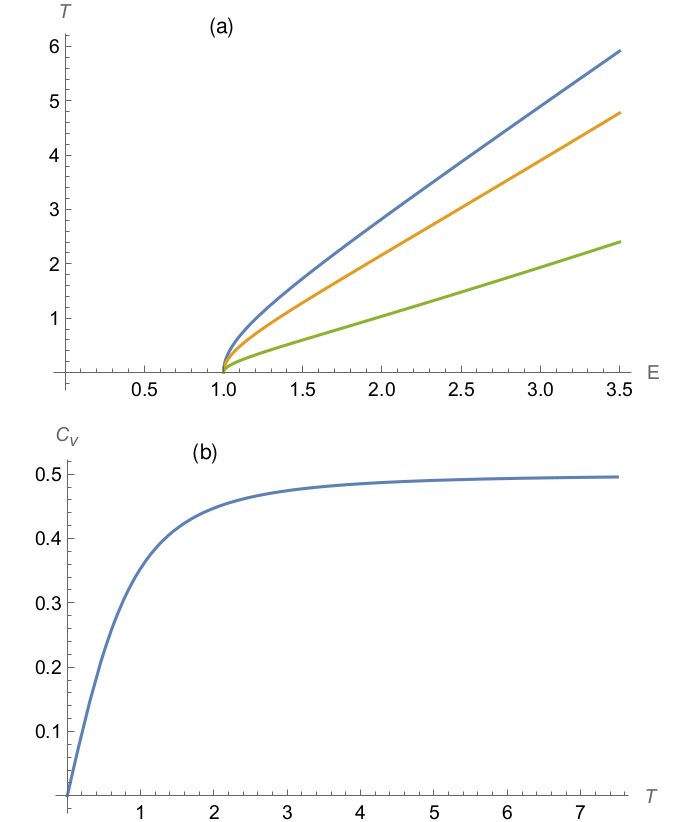}
\end{center}
\caption{Potential with central repelling step as a minimalist model of a real gas. a) Temperature-energy plot for the model (\ref{eq:steprealgas}), with $K=1$, $m=1$, $L=2$ for different exponents ($\alpha=0$ blue, $\alpha=0.5$ orange, $\alpha=0.9$ green). b) Heat capacity as a function of temperature for the case $\alpha=0$.} \label{fig:wall}
\end{figure}

We also considered the case of a smooth bump in the potential. This adiabatic  bouncer with parabolic repulsive potential about $x=0$ has the following Hamiltonian, 
\begin{equation}
H = \left\{ 
\begin{array}{lr}
\cfrac{p^2}{2 m} +\cfrac{k}{2} \Big(1 - x^2 \Big) & x \in (-1,1) \\[12pt]
 \cfrac{p^2}{2 m} & x \in [-L,-1] \cup [1,L].
\end{array}   
\right.
\end{equation}
For this system, Figure 4 shows the effect of a slow expansion of the right wall. Initially, the pressure is reduced, while entropy remains approximately constant by shrinking vertically to compensate for the increased horizontal dimension. However, at some point because of the reduction in temperature (i.e., average kinetic energy) the oscillator is not able to pass the bump and is repelled back to the right wall without reaching again the left wall. At this point the entropy (and of course the area enclosed by the trajectory) has a sudden drop marked by a discontinuous jump. A similar sudden drop in entropy is of course observed also in the previous case of the stepwise central bump in the potential. Such behaviors are reminiscent of a phase transition similar to a condensation by adiabatic expansion and cooling. While such behaviors are very interesting and reveal further valuable connections to real thermodynamic systems, analysis of phase transitions originating from the extended Helmholtz thermodynamics is left to future contributions.

\begin{figure}
\begin{center}
\includegraphics[width=212pt]{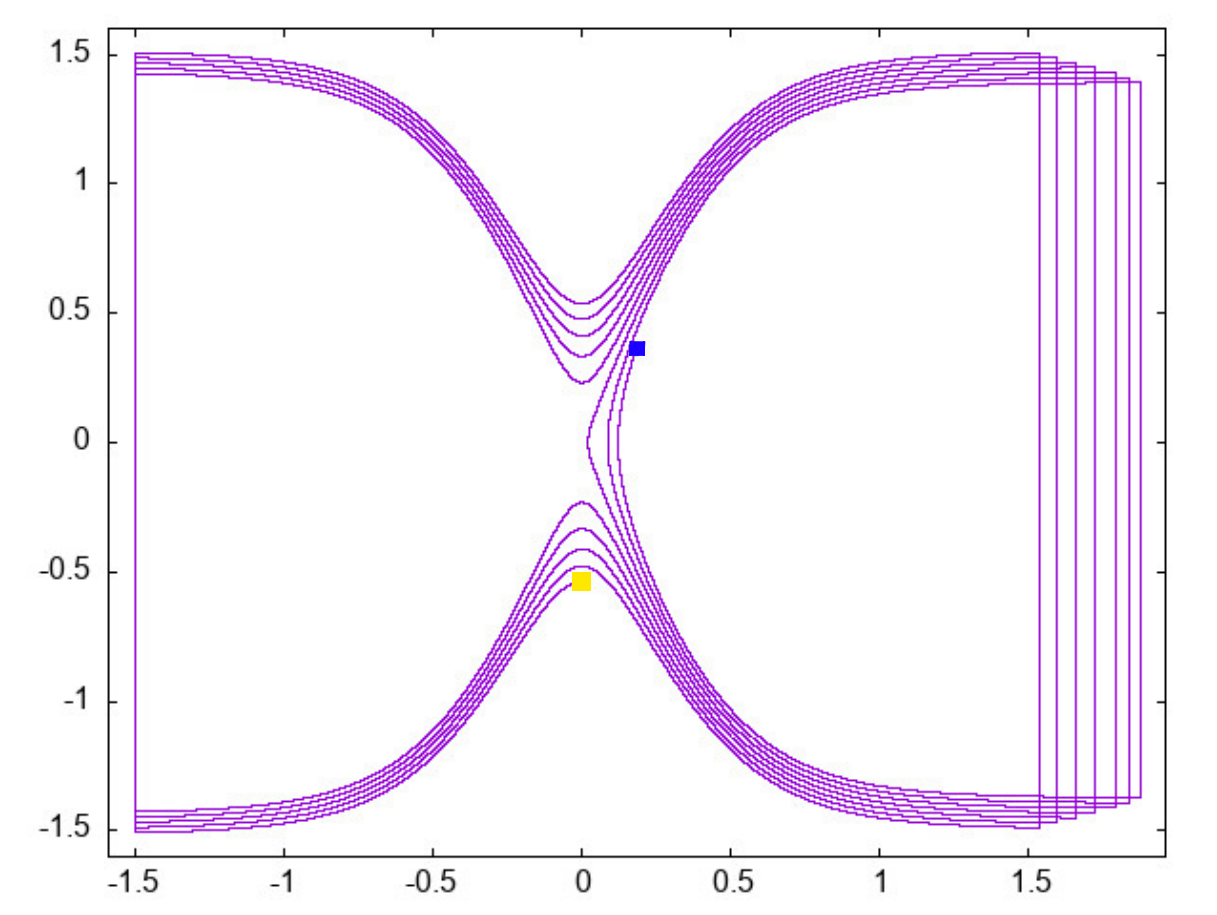}
\end{center}
\caption{Phase space trajectory of an adiabatic bouncer with parabolic obstacle and right wall moving like a piston expanding the volume of the gas. The right wall moves to the right with speed 0.01,
reducing the kinetic energy of the particle at each collision. The initial speed of the particle is 1. When the energy turns too small compared to the obstacle potential, the bouncer ``condensates'' in the right half of the volume. The region of the condensate is random, as it depends on the random initial condition. The initial condition is evidenced by a yellow square approximately at the center of the box. The final condition is in the confined region, evidenced by a blue square.} \label{fig:wall}
\end{figure}

\section{Systems in Thermal Contact with the Environment}
To extend Helmholtz theory beyond adiabatic conditions, in this section we introduce the coupling to an environment to allow for heat fluxes. 
In general, when the 1D oscillator is connected to an environment, the latter will provide some extra force, which may be generically expressed as:
\begin{equation}
\dot{x} = \cfrac{p}{m} \, , \quad 
\dot{p} = F - \alpha
\label{eqmot}
\end{equation}
where $\alpha$ depends on the environment and, in general, it makes the equations of motion non-autonomous. Given the initial condition 
$(q_0,p_0)$, lets us denote by
\begin{equation}
    x=x_\alpha(t;q_0,p_0) ~~ \mbox{and} ~~ p = p_\alpha(t;q_0,p_0),
\end{equation}
the corresponding solution of Eq.\ \eqref{eqmot}. The coupling with the environment has a different effect on the thermodynamic formalism depending on the time scales of the energy exchanges between the system and the environment. 
In molecular dynamics, a wide set of effective forces $\alpha$ has been developed, to express the effect of the environment on the system of interest, while various constraints fix the state point 
\cite{hoover1993nonequilibrium,hoover2012computational,rapaport2004art,j2007statistical,jepps2010deterministic,todd2017nonequilibrium}. It is not necessary, here, to discuss this quite important line of research, because our purpose is fulfilled when the effect of the environment is represented by any kind of external force.
Two limit cases are particularly interesting here. 
The first results from a fast timescale forcing giving rise to periodic oscillations with zero net exchange over a cycle. This does not induce any heat flux, because the thermodynamic formalism considers averaged conditions over a cycle. However, as we will see in the following examples, it modifies the fundamental equation and the related equations of state. The second case takes place when the coupling operates quasi-statically, that is on very long timescales: this slow-acting force alters the entropy of the system, producing a heat flux, $TdS$. We will discuss this extension considering a tunable thermostat in the special case of the elastic bouncer.

Additionally, some general considerations can be made revisiting the virial and adopting the equipartition expression for the average kinetic energy, which gives 
\begin{equation}
    P = \cfrac{1}{L} \, k_{_B} T + \cfrac{1}{L} \, \mathcal{V} - \cfrac{\overline{\alpha q} }{L}, 
\end{equation}

modifying the previous equations of state in a way that depends on the details of the interaction $\alpha$. Suppose system (s) and environment (e) form together a Hamiltonian system, depending on various parameters $\lambda_s$, $\lambda_e$ and $\lambda_{se}$. Denoting by $(x_s,p_s)$ the coordinates and momenta of s, by $(x_e,p_e)$ the coordinates and momenta of e, and by $\Gamma$ the whole set, the corresponding Hamiltonian may be assumed to take the following form:
\begin{equation}
    {\cal H}(\Gamma; \lambda_s , \lambda_e , \lambda_{se}) = 
    H_s(x_s,p_s; \lambda_s, \lambda_e , \lambda_{se}) +
    H_e(x_e,p_e; \lambda_e) +
    H_{es}(\Gamma; \lambda_{se}) 
\end{equation}
where each parameter my be replaced by vectors of parameters, for sake of completeness. The corresponding equations of motion of the system take the form:
\begin{equation}
    \dot{x}_s = \cfrac{\partial H_s}{\partial p_s} + \cfrac{\partial H_{se}}{\partial p_s} = \cfrac{p}{m} + g^{se}\, ; \quad 
    \dot{p} = - \cfrac{\partial H_s}{\partial x_s} - \cfrac{\partial H_{se}}{\partial x_s} = F + F^{se}
\end{equation}
where $F$ may be decomposed in the usual wall (if any) and interaction forces, and $F^{se}$ and $g^{se}$ and $F^{se}$ depend on the interaction between s and e, on which various hypotheses have been advanced, most notably by Kirkwood, within the theory of dense fluids \cite{kirkwood1935statistical,jarzynski2017stochastic}. The equation of state would then take the form:
\begin{equation}
    P = \cfrac{1}{L} \, k_{_B} T + \cfrac{1}{L} \, \mathcal{V} - \cfrac{\overline{p g^{se}} + \overline{q F^{se}}}{L} 
\end{equation}
where the last fraction needs to be related to a particular physical situation, in order to be interpreted. Together with the interaction term ${\cal V}/L$, it constitutes the counterpart of the virial expansion for standard particle systems. 

In turn, the fundamental relation takes the form
\begin{equation}
\overline{E_s}(\lambda_s, \lambda_e , \lambda_{se}) =
    \overline{E}(\lambda_s , \lambda_e , \lambda_{se}) -     \overline{E_e}(\lambda_e) -
    \overline{E_{se}}(\lambda_{se}) 
\end{equation}
where $\overline{E}_s$ may be interpreted as the internal energy of s, 
$\overline E$ the energy of s and e together, $\overline{E}_e$ that of e alone, and $\overline{E}_{se}$ that of the interaction of s and e.
Under various hypotheses, and for specific models, such a relation can be differentiated with respect to the parameters (which may include some of the energies in it) and Legendre transformed to produce new fundamental representations or material properties.

\subsection{Isokinetic systems}
One of the popular ways of coupling an autonomous Hamiltonian system to an environment is afforded by the Gaussian isokinetic thermostat \cite{j2007statistical}. 
In this case, the kinetic energy is preserved in time, $K=\overline{K}$, while the total and the  potential energies are allowed to vary in time, and their time averages, $\overline{E}$ and $\overline{\Phi}$, are then determined by the form of the equations of motion and by the value of $K$, which obviously equals its time average $\overline{K}$. The corresponding equations of motion are the following:
\begin{equation}
 \dot{x} = \cfrac{p}{m} \, \, , ~~ 
 \dot{p} = - \frac{\partial \Phi}{\partial x} - \alpha(x,p) p \, \, , ~~ \alpha = \cfrac{F \cdot p}{p^2} \, \, , ~~ 
 F = - \frac{\partial \Phi}{\partial x},
 \label{IK1d}
\end{equation}
where, $F \cdot p$ represents a scalar product, and reminds us that all terms could be vectors and concern more than one particle in more than one dimension. In the 1-dimensional case, Eq. \eqref{IK1d} turns:
\begin{equation}
 \dot{x} = \cfrac{p}{m} \, \, , ~~ 
 \dot{p} = F -  \cfrac{F p}{p^2} \,  p = 0
 \label{IK1d1}
\end{equation}
apparently like the free particle, that  moves with fixed speed 
$|v| = \sqrt{2 K / m}$. If this particle is confined between hard walls, the motion is periodic, and the total and potential energies obey:
\begin{equation}
E(t) = K + \Phi(x(t; x_0 , v_0) ) \, , \quad
\overline{E}(K) = K + \cfrac{1}{\tau} \int_0^\tau 
\Phi(x(t; x_0 , v_0) )  d t \, , \quad \overline{\Phi}(K) = \cfrac{1}{\tau} \int_0^\tau 
\Phi(x(t; x_0 , v_0) )   d t,
\end{equation}
where $(x_0,v_0)$ is an initial condition  such that $mv_0^2/2=K$.
As in the case of the elastic bouncer, 
the infinitely hard barriers do not appear in the equations of motion; they only act  at the ends of the segment $[0,L]$, where they reverse the direction of motion without slowing it down. 

This strongly idealized model of a system in contact with a heat bath, meant to be isothermal, suffices to grasp the main concepts. In this case, identifying $2K$ with $k_{_B}T$, the virial leads to the perfect gas law:
\begin{equation}
    P = \cfrac{1}{L} \, k_{_B} T
\end{equation}
because only the term $-\overline{x F_L^{\rm wall}}/2$ contributes to the kinetic energy, since $\dot{p}=0$.
This is correct, although the total energy is not merely kinetic or dominated by the kinetic energy, and the average potential $\overline \Phi$ does not need to be negligible. 
For instance, in the case of harmonic potential $\Phi_k = k x^2 / 2$, and hard walls at 0 and $L$, we have:
\begin{equation}
    \cfrac{\tau}{2} = L \sqrt{\cfrac{m}{2K}} \, , \quad
    \overline{\Phi_k} =
    \cfrac{2}{\tau}
\int_0^{\tau/2} \cfrac{k}{2} ~ 
x(t)^2 ~ d t = 
    \cfrac{2}{\tau}
\int_0^{\tau/2} \cfrac{k K}{m} ~t^2 ~ d t = \cfrac{k L}{3}
\end{equation}
irrespective of the value of $K$, because it does not matter how fast the distance is travelled: speed being uniform, all values of the potential are equally weighted. 
It therefore constitutes a system energetically different from an elastic bouncer subject to no external or internal fields, that merely moves back and forth between two hard walls. The differences depend on the form of $\Phi$, and while the equation of state does not change with $\Phi$, the thermodynamic fundamental relation does. In particular, if $\Phi=\Phi_\lambda$ depends on a parameter $\lambda$, and we set $2 K = k_{_B} T$, in the energy representation we have:
\begin{equation}
\overline{E} = \overline{E}(T,L,\lambda) =
\cfrac{k_{_B} T}{2} + \overline{\Phi_\lambda}(T,L,\lambda)
\end{equation}
For the harmonic potential $\Phi_k$,
one obtains:
\begin{equation}
\overline{E} = \overline{E}(T,L,k) =
\cfrac{k_{_B} T}{2} + \cfrac{k L}{3}
\, ; \quad 
C_{v,k} = \left. \cfrac{d \overline{E}}{d T} \, \right|_{L,k} = \cfrac{k_{_B}}{2} 
\end{equation}
as for the elastic bouncer, because the potential term does not depend on $T$.

\subsection{Periodically Driven System}

A periodic driving also corresponds to coupling to an environment. The equations of motion in general are: 
\begin{equation}
    \dot{x} = \cfrac{p}{m} \, , \quad
    \dot{p} = - \frac{\partial \Phi}{\partial x} 
     + m A \cos{ \omega t}
\end{equation}
where $\Phi$ can take different forms, but it suffices to investigate the harmonic oscillator with frequency $\omega_0$. 
The solution of the corresponding equations of motion is then given by:
\begin{equation}
    x(t) = \cfrac{A}{\omega_0^2 - \omega^2} \, \cos{\omega t} + B \cos\left( \omega_0 t + \varphi \right) \, ; \quad
    \dot{x}(t) = - \cfrac{A \omega} {\omega_0^2 - \omega^2} \, \sin{\omega t}
- B \omega_0 \sin \left( \omega_0 t + \varphi \right) 
\end{equation}
where $B$ and $\varphi$ are determined by the initial conditions.
Note that this can be directly obtained projecting to the space of the particle of interest the dynamics of a system with two oscillators, the second of which plays the role of a very massive environment e, which is unaffected by the motion of the system. The equations of motion can then be written as:
\begin{equation}
    \dot{x} = \cfrac{p}{m} \, , \quad
    \dot{p} = - \frac{\partial \Phi}{\partial x} 
     + m \theta \, , \quad
     \dot{\theta} = \phi \, , \quad \dot{\phi} = -\omega^2 \theta
\end{equation}
where the initial conditions set the constants.
There are two cases: a) $\omega / \omega_0$ is rational, hence the dynamics are periodic with the minimum common multiple of the two periods, $\tau = 2\pi/\omega$ and $\tau_0=2\pi/\omega_0$, as period; b) the motion is only quasi-periodic. In the first case, the time averages can be computed integrating over the period, in the second case, one must average over all positive times. Begin with $\omega = n \omega_0$, hence $\tau = \tau_0/n$, where $n$ is a positive integer. The energy fluctuates, and its average is given by:
\begin{equation}
    \overline{E}(A,E^0,\omega_0,\omega) = E^0
    + \cfrac{(1 + n^2) k A^2}{4 (1 - n^2)^2 \omega_0^4} \, , ~~ \mbox{where } ~~
    \overline{K} = \overline{K}^0 + 
    \cfrac{k n^2 A^2}{4 \omega_0^4 (1-n^2)^2} ~ , ~~
    \overline{\Phi} = \overline{\Phi}^0 + 
    \cfrac{k A^2}{4 \omega_0^4 (1-n^2)^2} 
\end{equation}
which constitute the 0-pressure fundamental relation.
As in the case of the pure harmonic oscillator, the pressure vanishes, if
$L \ge \ell$, while it is positive if $L < \ell$. Compared to the purely harmonic oscillator, the isolated (0 pressure) oscillations amplitude and period change. Moreover, given the new $\ell$ and $\tau$, 
the interaction term is now supplemented with a non potential external force, and we have 
\begin{equation}
\overline{x F} = -k \overline{x^2} +
m A ~ \overline{\, x \cos{\omega t} \, } = 
{\cal V}(L,A,B,m,k,\omega,\omega_0,\varphi) 
\end{equation}
apart from the usual hard wall term. The explicit expression, in terms of average energy and kinetic energy, will be obtained inverting the dependence on the initial conditions, that determine $B$ and $\varphi$, once the other parameters are fixed. A similar situation is obtained if $\omega_0 / \omega$ is a generic rational number although $\ell$, computed as in Eq. \eqref{elle}, 
is not the distance between the two ends of the oscillations. It is the length of the 
trajectory, possibly traversed going back and forth various times before reaching the extremes, so in general it is longer than 
$(x^+ - x^-)$. What matters for a positive pressure, even in this case, is that $L < (x^+ - x^-)$. It should be noted, that the phase $\varphi$ will shift 
at every collision with the walls, but it shall be a function of the initial phase and of $L$.
In general, adopting the usual equality 
$2\overline{K}=k_{_B} T$, one still has:
\begin{equation}
    P = \cfrac{1}{L} ~ k_{_B} T + 
    \cfrac{1}{L} ~ \cal V
\end{equation}
where $\cal V$ may be rather involved as a function of the relevant average quantities and of $L$. The fundamental relation, can be analogously obtained, averaging the sum of potential and kinetic energies:
\begin{equation}
     \overline{E} = \overline{E} (L,T,\mbox{parameters}).
\end{equation}

\begin{figure}
\begin{center}
\includegraphics[width=208pt]{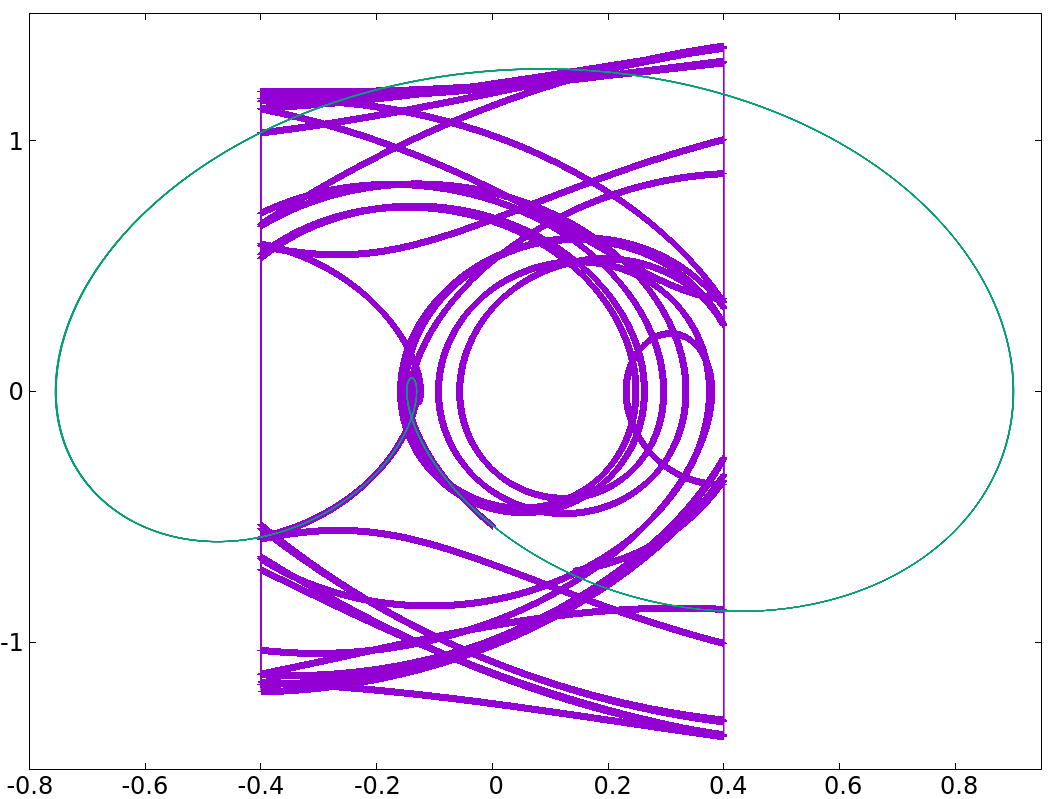} ~~~
\includegraphics[width=208pt] {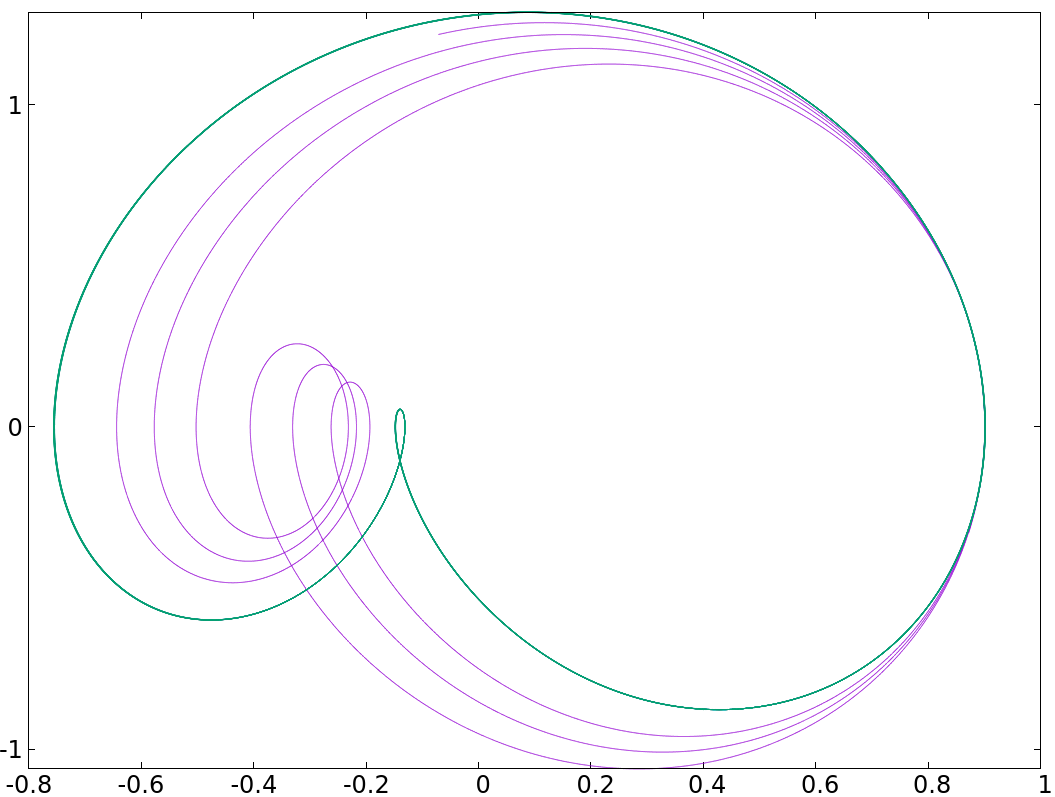}
\end{center}
\caption{Harmonically forced oscillator and adiabatic standing walls.
Left: thin green line draws a kind of heart, extending from about -0.8 to about 0.9, while the thick purple line shows what happens if two walls are introduced at a distance quite shorter than the width of the heart, at -0.4 and 0.4. 
Right: green line is the heart, purple line is an initial transient when the walls are at -0.9 which is outside the heart, and 0.9 which slightly touches the right end of the heart. Initially, the heart almost preserves its shape, but smaller. Then, collisions with the wall make it slowly increase the oscillator energy, which leads to a slow growth of the heart.}
\label{heart1}
\end{figure}

\begin{figure}
\begin{center}
\includegraphics[width=208pt]{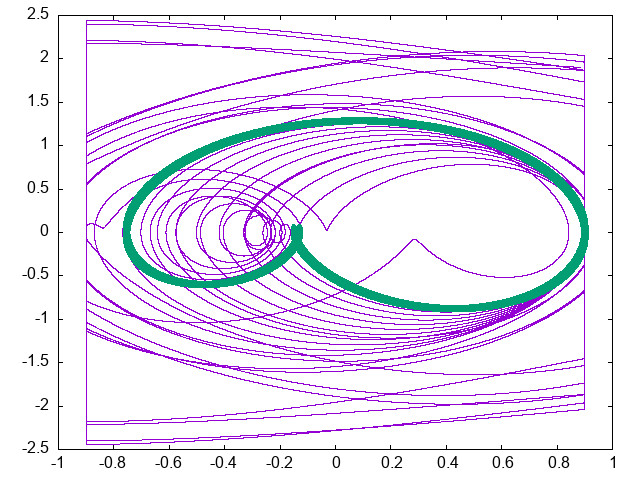} ~~~
\includegraphics[width=208pt]{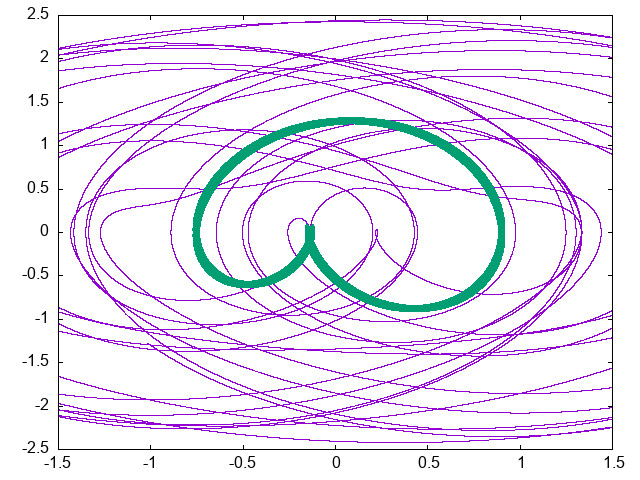}
\end{center}
\caption{Harmonically forced oscillator and adiabatic standing walls.
Left: thick green line shows the heart with no walls, extending from about -0.8 to about 0.9, while the thin purple line shows the steady state that follows the right panel of Fig.\ \ref{heart1}, when the growing transient hearts collide with the left wall at -0.9. 
Right: when the forcing is almost resonant (forcing frequency 1.1) with the natural frequency (1.0) of the harmonic oscillator, and walls are placed -1.5 and 1.5, {\em i.e.}\ do not touch the heart, the amplitude of the oscillations grows till the walls are reached. A non-periodc stationary state seems to be reached. At resonance, it is not clear whether a slow growth lasts forever, or the walls, confining the amplitde of the positions, also limit the momenta, hence the energy.}
\end{figure}

\subsection{Heat Fluxes in a Thermostatted Elastic Bouncer}

Consider a particle subject to no forces, except those exerted by the walls. If the walls move, expanding or compressing the volume, the bouncer's energy decreases or increases, respectively. To maintain the kinetic energy constant, one may introduce the following thermostatted equations of motion:
\begin{eqnarray}
&& \dot{x} = v \\
&& \dot{v} = \cfrac{1}{\tau}
\, \cfrac{v_R^2 - v^2}{2 v_R v}
\label{eq:lambthermo}
\end{eqnarray}
where $v_R$ is the reference speed, corresponding to the chosen temperture and $\tau$ is the characteristic relaxation time. Unlike the Gaussian thermostat, these equations of motion do not fix the kinetic energy, but apply a restoring force, toward the target kinetic energy. The analytic solution, for a free flight from collision with a wall to the next is given by:
\begin{equation}
    v = \pm \sqrt{v_R^2 + r e^{-t / \tau v_R^2}}
\end{equation}
where the sign and $r$ are determined by the initial condition, or by the starting point immediately after a collision. For speed  $|v(0)|$ smaller than $v_R$, $r$ is negative; for $|v(0)|$ larger than $v_R$, $r$ is positive.  If walls move, energy at collisions with them changes, and the thermostat tries to restore the target kinetic energy.
\begin{figure}
\begin{center}
\includegraphics[width=208pt]{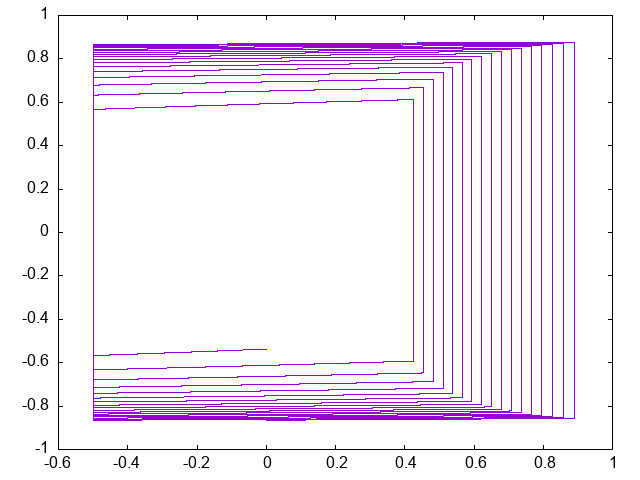} ~~~
\includegraphics[width=208pt]{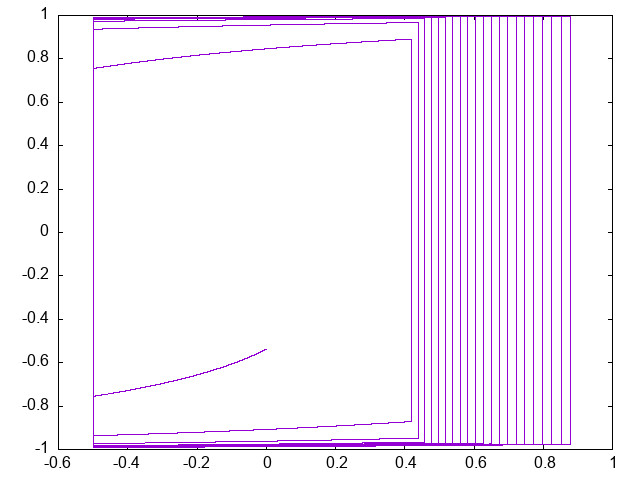}
\end{center}
\caption{Thermostatted bouncer.
Left: the process is too fast for the relaxation time of the thermostat: the right wall acts like a piston rapidly
expanding the volume of a gas. Adjustment of kinetic energy is not efficient.
Right: the characteristic time of the thermostat is sufficiently short, compared to the piston speed and the picture in phase space is almost that of expanding rectangles with fixed height.}
\end{figure}

\begin{figure}
\begin{center}
\includegraphics[width=208pt]{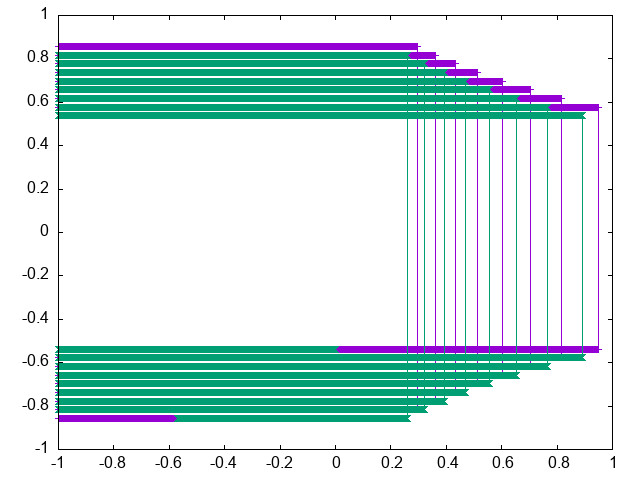} ~~~
\includegraphics[width=208pt]{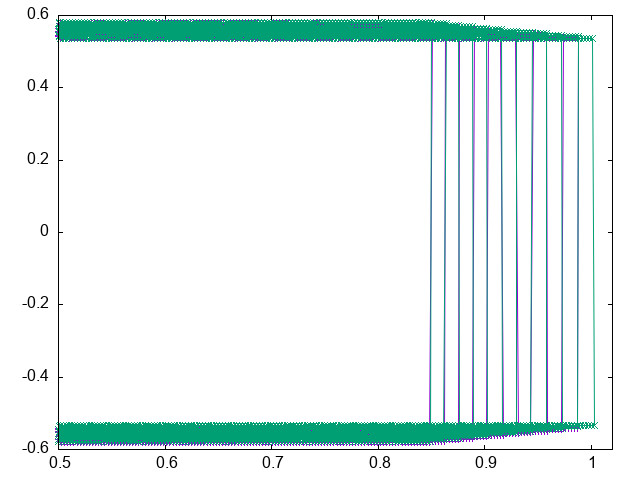}
\end{center}
\caption{Isolated bouncer (adiabatic walls).
Left: purple lines refer to the case in which the} right wall acts like a piston compressing a gas, and moves fast (speed 0.02). Green curves refer to the reverse motion of the piston. Under compression, the kinetic energy grows: cf.\ the growing values of the purple horizontal lines, while the purple vertical lines, representing the position of the piston, move to the left.  This process is not macroscopically reversible. The microscopic initial condition matters: inverting the motion of the piston without inverting the velocity of the particle (green lines), most momenta are the same (horizontal lines overlap), but positions are not traced back: 
the green vertical lines do not coincide with the purple ones. 
Right: slow piston (speed 0.002) and positions larger than 0.5 are portrayed. The energy variation is reduced 
(purple and green horizontal lines overlap in a thin strip), and the macroscopic reversibility of the process is evidenced
by the overlap of the horizontal and the vertical lines (the microscopic initial condition does not matter).
\end{figure}

\section{Carnot Cycle with the Elastic Bouncer}

We can now join quasi-static transformations on oscillators to perform thermodynamic cycles and use them as thermodynamic engines to do work using heat flows. To this goal, we consider the simplest case of an elastic bouncer on which we perform the classical sequence of the Carnot cycle: isothermal and adiabatic expansions followed by isothermal and adiabatic compression. For the adiabatic expansion/compression we can consider that, because of adiabatic invariance, 

$I=const$ and Eq.\ \eqref{BouHJ} implies $H \propto L^{-2}$ and $v\propto L^{-1}$. Thus, using (\ref{eq:pressbouncer}), it follows that for adiabatic transformations, the pressure goes as $P\propto L^{-3}$.

For the isothermal transformations, the kinetic energy does not change and therefore also $v$ does not change. Thus, based on (\ref{eq:pressbouncer}), $P\propto L^{-1}$. Practically, for a quasi-static, isothermal transformation, in order to keep the velocity constant, we can give an infinitesimal kick to the particle immediately after each collision with the wall, so that the velocity changes as $dv=-1/L^2dL$. 
Mathematically this can be achieved

taking the thermostat (\ref{eq:lambthermo}) in the limit of an infinitesimal relaxation time.
It is clear that when plotted in the $PV$ plane, one gets the typical shape of the Carnot cycles, in this case bounded by the intersections of the $L^{-3}$ power law of the adiabatic and the hyperbola $L^{-1}$ of the isothermal transformation. In the $ST$ plane one obtains the classical rectangular shape of the cycle,
given by the sequence of isothermal and isentropic (adiabatic) transformations. 

\begin{figure}[htp]
\begin{center}
\includegraphics[width=208pt]{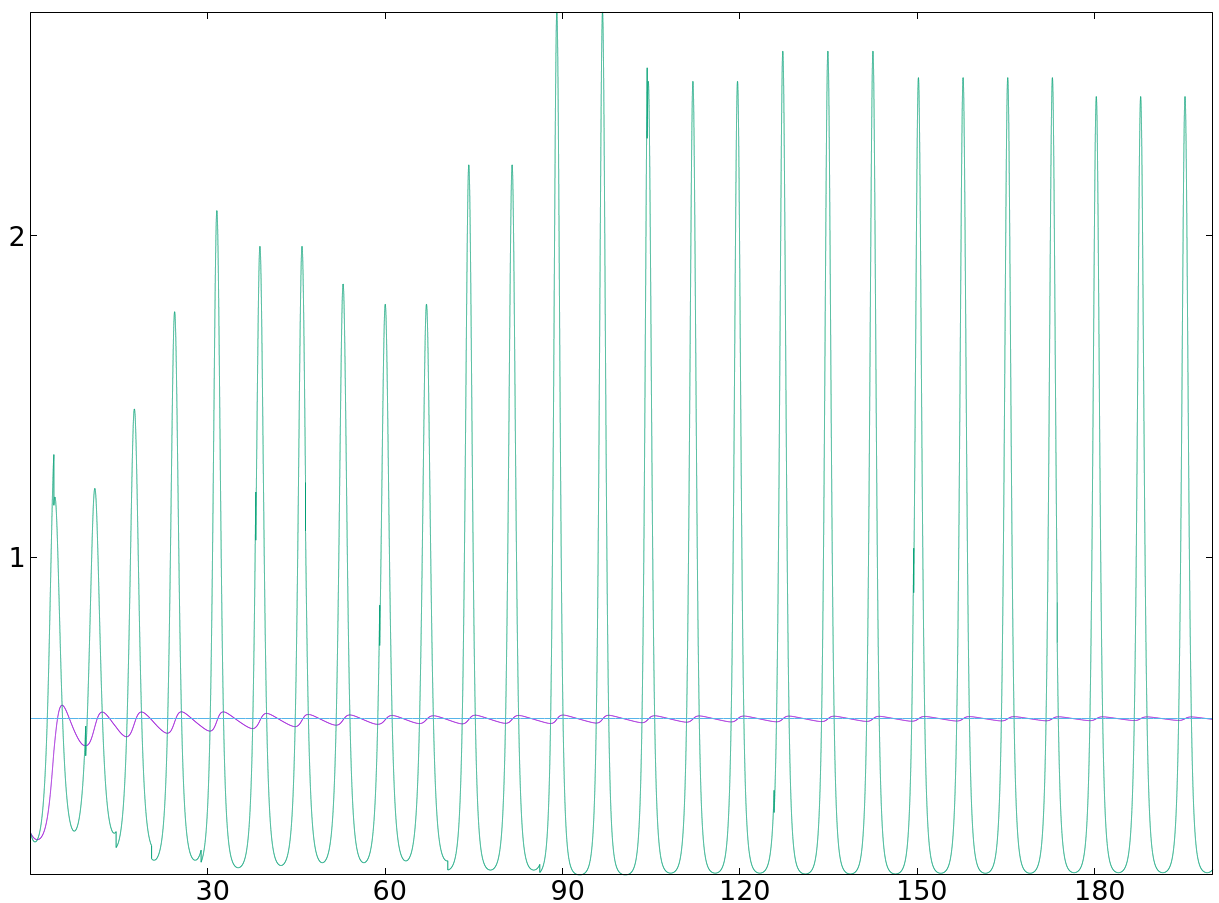} ~~~
\includegraphics[width=208pt]{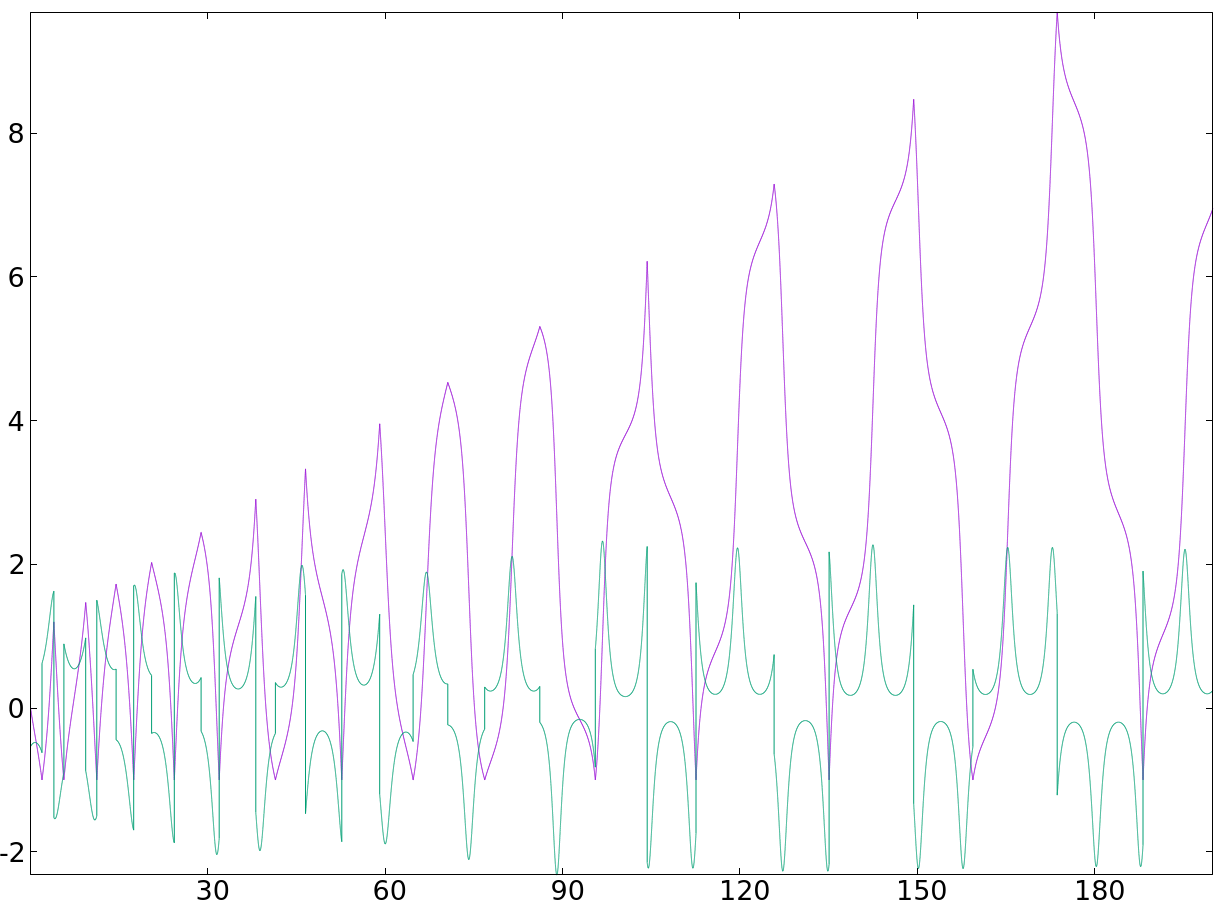}
\end{center}
\caption{Isothermal expansion with the Nose-Hoover thermostat. Left:
time average (purple curve), instantaneous values (green curve) and target value (cyan horizontal line) of the 
kinetic energy. Right: instantaneous momenta (green curve) and positions (purple curve). 
The thermostat is fast to adjust the temperature.
Left wall is fixed and the speed of right wall is 0.05.
The position of the moving wall is indicated by the peaks of the purple curve in the right panel.} 
\label{NHth}
\end{figure}

We also performed a finite time Carnot cycle by considering numerically the elastic bouncer with moving walls. For the isothermal transformation we considered different options for the thermostat: 1) Gaussian IK thermostat, for which the bouncer speed is always 1; 2) Nos\'e-Hoover thermostat \cite{hoover1993nonequilibrium}, which lets the speed fluctuate about the chosen mean; 3) Eq.\ (\ref{eq:lambthermo}), which exponentially in time drives toward the chosen speed. 
While the first is obvious, the case of NH is more interesting. An example of an isothermal expansion for the NH thermostat is presented in Fig.\ \ref{NHth}. 
During the expansion the kinetic energy varies quite widely 
(green curve of left panel of Fig.\ \ref{NHth}), but its averaged value over a period of oscillation 
(purple line of left panel of Fig.\ \ref{NHth})
is stable and, after a short transient, settles on the value corresponding to the desired temperature. The strong kinetic energy fluctuations, however, make more difficult to use this thermostat, because to switch to the adiabatic expansion, needed for the Carnot cycle, one would have wait for the exact moment in which the energy is the desired one. The right panel of Fig.\ \ref{NHth} portrays the 
instantaneous momenta (green line) and positions (purple line) for this case, showing the rapid growth of the volume available to the bouncer. 

\begin{figure}[htp]
\begin{center}
\includegraphics[width=208pt]{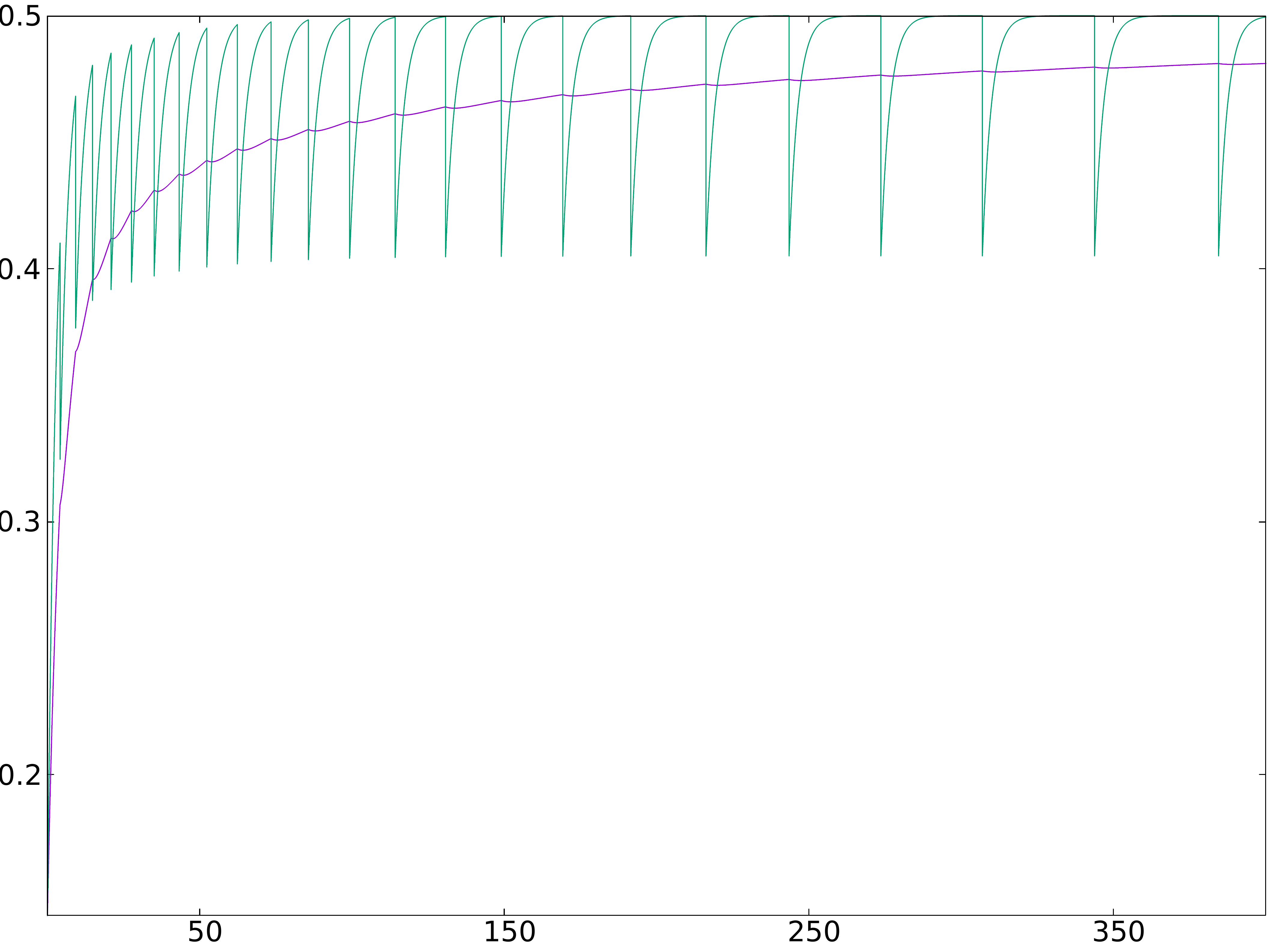} ~~~
\includegraphics[width=208pt]{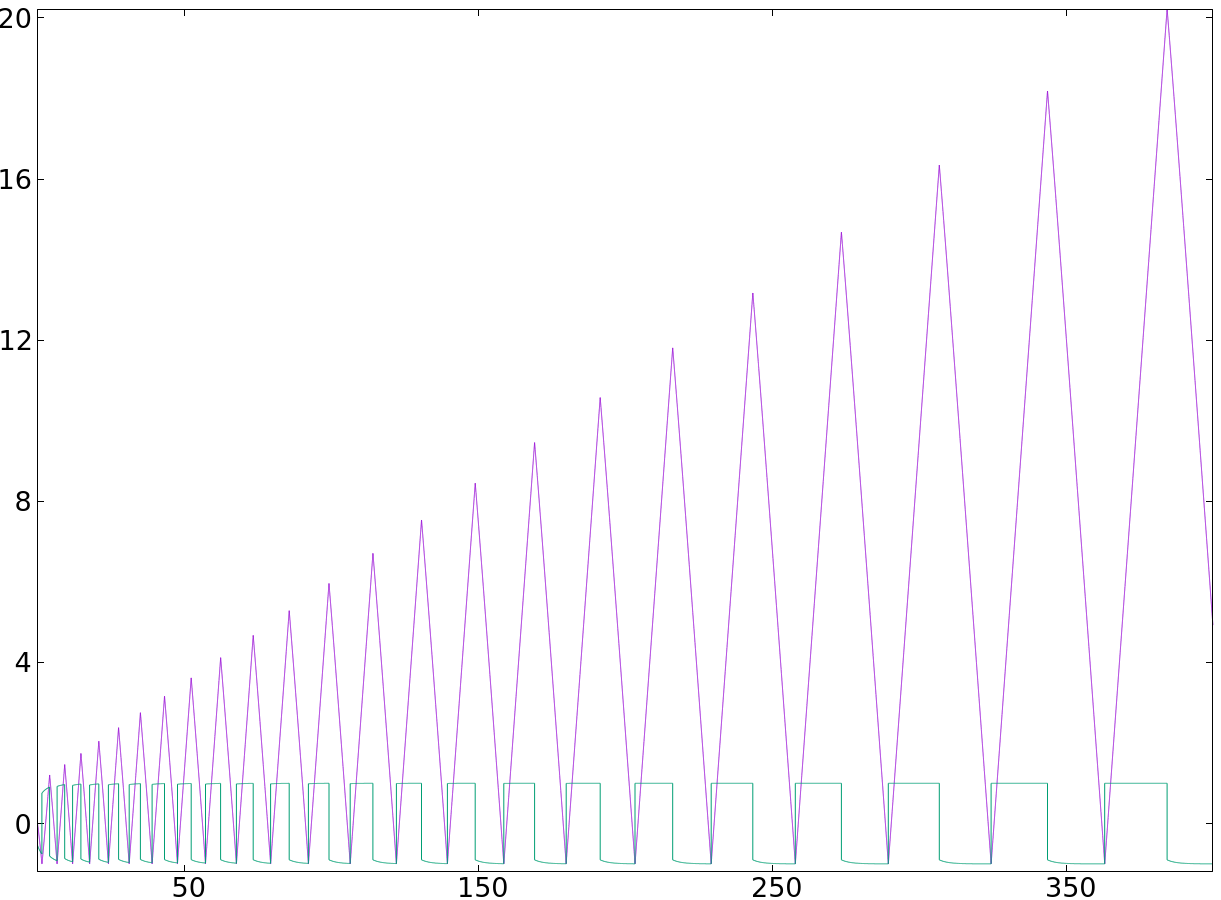}
\end{center}
\caption{Isothermal expansion with exponential thermostat (\ref{eq:lambthermo}). 
Left: time average(purple line) and instantaneous values (green line) of the kinetic energy. At each collision with the expanding wall, the energy decreases, then the thermostat drives it back to the target value. As the volume increases, such fluctuations are less frequent, and the target value more stable. Right: Instantaneous positions
(purple line) and momenta (green line). The peaks of the purple line reveal the position of the moving wall.} Left wall is fixed, speed of right wall is 0.05. The thermostat is relatively fast to adjust the temperature.
\label{ExpThPer}
\end{figure}

For the exponential thermostat, the corresponding figure is presented in Fig.\ \ref{ExpThPer},
for a slow thermostat (large $\tau$), to test its performance in disadvantaged conditions. While the average kinetic energy adjustment times are slower 
(purple line in the right panel of Fig.\ \ref{ExpThPer})
, the figure also shows that the instantaneous kinetic energy fluctuations are  smaller and rapidly reach the target value
(green line in the right panel of Fig.\ \ref{ExpThPer}). This makes the switch to the subsequent adiabatic transformation much easier. 
For this reason,
and because conceptually nothing changes,
we only employ the exponential thermostat (\ref{eq:lambthermo}) to perform the entire sequence of the Carnot cycle. 

\begin{figure}[htp]
\begin{center}
\includegraphics[width=190pt]{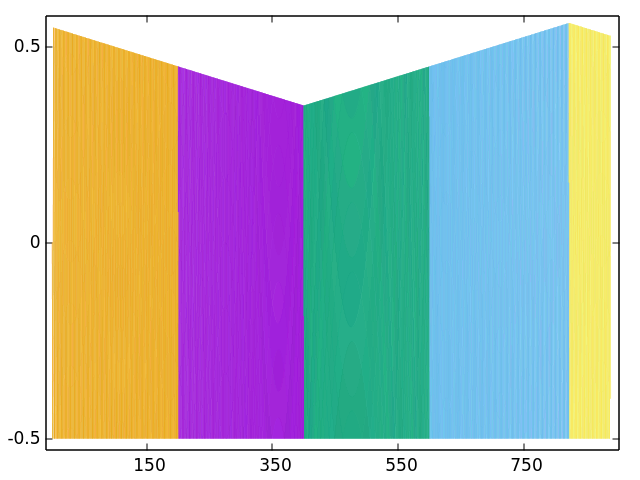}
\includegraphics[width=190pt]{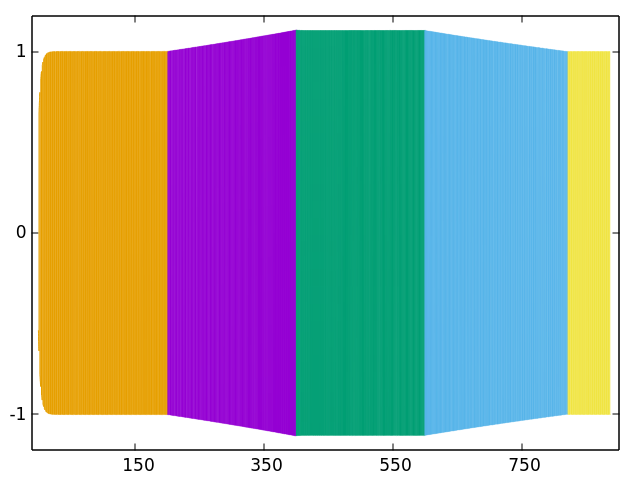} \\
\includegraphics[width=190pt]{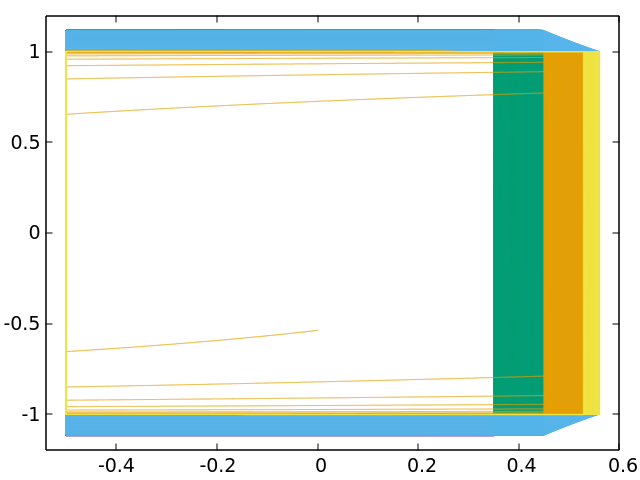}
\includegraphics[width=190pt]{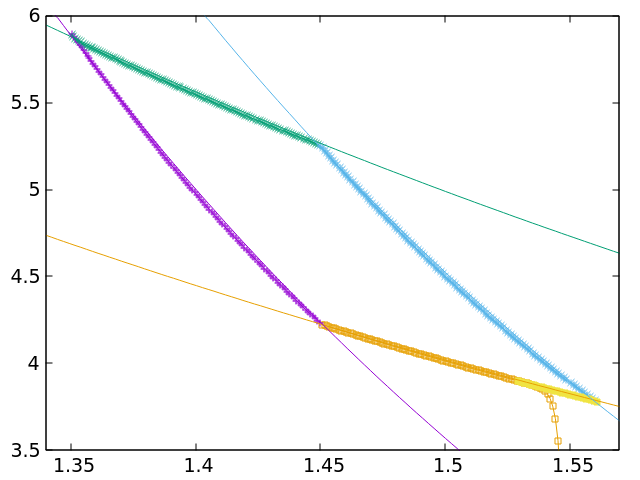}
\end{center}
\caption{Carnot cycle with exponential thermostat. Left wall is fixed, speed of right wall is 0.001. In every step of the cycle, the bouncer goes back and forth many times, approximating a quasi-static transformation.
Top left: particle position as a function of time. Top right: momentum as a function of time. The isothermal compression (orange) begins with a transient in which the kinetic temperature equilibrates with the thermostat. The adiabatic compression (purple),  isothermal expansion (green), and adiabatic expansion (cyan) follow. The final yellow lines represent the isothermal compression, that closes the cycle, when  transients have decayed. Bottom left:
phase space trajectory: isothermal compression (orange and yellow); adiabatic compression (hidden by later transformations); isothermal expansion (green); adiabatic expansion (cyan). 
Bottom right: Carnot cycle in the PV plane.
The transient appears as the almost vertical orange line, that becomes isothermal compression. Adiabatic compression (purple), isothermal expansion (green) and adiabatic expansion (cyan) follow. The following isothermal compression (yellow) closes the loop.
}
\label{CarnFig}
\end{figure}

The results for the Carnot cycle are shown in Figure \ref{CarnFig}. In these simulations, the speed of the wall is kept quite slow, to approximate quasi-static transformations. Also shown is the short initial transient, for the isothermal expansion, in which the speed is rapidly adjusted to its target value.

The top left panel of this figure shows the many flights between the two walls that the bouncer experiences in time, as the right wall slowly moves. Orange corresponds to isothermal compression, starting from a non-equilibrated state which rapidly equilibrates; purple to adiabatic compression; green to isothermal expansion; cyan to adiabatic expansion; yellow to a piece of the isothermal compression to close the loop. All panels adopt the same color coding. The top right panel shows the momentum oscillations. The orange line reveals that the initial transient is rapidly quenched, and then it matches with the final compression depicted by the yellow line. The purple line shows how momentum increases as the wall compresses the bouncer; the green shows that in the isothermal expansion the momentum size is fixed; the cyan line shows the loss of momentum when the volume expands. The bottom left panel represents the phase space trajectory. The orange line show the rapid decay of the transient, as momentum starts small compared to the isothermal target. It is interesting to see how this happens. The purple line representing the adiabatic compression is behind the isothermal and the adiabatic expansion, consistently with the top panels. The thickness of the horizontal lines corresponds to the variations of momentum in the adiabatic phases; the thickness of the vertical lines to the variations of volume. The bottom right panel eventually shows the full cycle in the PV plane, with the initial transient represented by the almost vertical  orange line with square symbols. The thick lines correspond to the simulations. The reversibility of the cycle is also illustrated by the perfect match of the initial and final isothermal compressions, when the transient has decayed.
The overall simulation approximates very well the Carnot cycle predicted theoretically for the quasistatic case, i.e., $L^{-3}$ for the adiabatic and $L^{-1}$ for the isothermal transformations, 
which correspond to the thin lines. The purely mechanical Carnot cycle thus obtained should be amenable to simple experimental verification, adding an interesting connotations to this extension of Helmholtz thermodynamics.

\section{Conclusions}

We considered an extension of Helmholtz thermodynamics for Hamiltonian oscillators to include non zero pressure, thanks to 
hard walls, and heat transfer from a coupling with the environment. This has allowed us to perform actual thermodynamic transformations, which in turn enabled us to obtain realistic thermodynamics with simple mechanical systems. 
Uncomplicated changes to the oscillator's equations have obtained behaviors that reproduce ideal gases and that display phase transitions. Interesting applications have demonstrated the potential of the extended approach, including a minimalist version of the Carnot cycle obtained as a sequences of transformations on the  oscillations of a simple elastic bouncer. 
This has also cast in plain terms the need for thermodynamic quantities to 
refer to equilibrium or very slowly evolving states. As they are defined by time averages, a fast change of the environment makes the microscopic initial conditions relevant for the average results. Slow variations, and the recurrent behaviour of the observables makes the averages involve a large number number of cycles, in which the microscopically defined quantities explore the same range of values. The averages then negligibly depend on the exact microscopic initial values. Then, the macroscopic quantities are sufficiently stable that the 
thermodynamic relations and balance equations make sense, and can be framed within regular mathematical expressions.

We believe there are several exciting directions for future research.
Several questions touch the foundations of thermodynamics, regarding, for instance, the  meaning of entropy changes.
Also, the obvious but nontrivial endeavor of extending the framework to multi-particle systems,
without recourse to non-mechanical concepts.


\bibliography{references}  

\begin{thebibliography}{28}%
\makeatletter
\providecommand \@ifxundefined [1]{%
 \@ifx{#1\undefined}
}%
\providecommand \@ifnum [1]{%
 \ifnum #1\expandafter \@firstoftwo
 \else \expandafter \@secondoftwo
 \fi
}%
\providecommand \@ifx [1]{%
 \ifx #1\expandafter \@firstoftwo
 \else \expandafter \@secondoftwo
 \fi
}%
\providecommand \natexlab [1]{#1}%
\providecommand \enquote  [1]{``#1''}%
\providecommand \bibnamefont  [1]{#1}%
\providecommand \bibfnamefont [1]{#1}%
\providecommand \citenamefont [1]{#1}%
\providecommand \href@noop [0]{\@secondoftwo}%
\providecommand \href [0]{\begingroup \@sanitize@url \@href}%
\providecommand \@href[1]{\@@startlink{#1}\@@href}%
\providecommand \@@href[1]{\endgroup#1\@@endlink}%
\providecommand \@sanitize@url [0]{\catcode `\\12\catcode `\$12\catcode
  `\&12\catcode `\#12\catcode `\^12\catcode `\_12\catcode `\%12\relax}%
\providecommand \@@startlink[1]{}%
\providecommand \@@endlink[0]{}%
\providecommand \url  [0]{\begingroup\@sanitize@url \@url }%
\providecommand \@url [1]{\endgroup\@href {#1}{\urlprefix }}%
\providecommand \urlprefix  [0]{URL }%
\providecommand \Eprint [0]{\href }%
\providecommand \doibase [0]{https://doi.org/}%
\providecommand \selectlanguage [0]{\@gobble}%
\providecommand \bibinfo  [0]{\@secondoftwo}%
\providecommand \bibfield  [0]{\@secondoftwo}%
\providecommand \translation [1]{[#1]}%
\providecommand \BibitemOpen [0]{}%
\providecommand \bibitemStop [0]{}%
\providecommand \bibitemNoStop [0]{.\EOS\space}%
\providecommand \EOS [0]{\spacefactor3000\relax}%
\providecommand \BibitemShut  [1]{\csname bibitem#1\endcsname}%
\let\auto@bib@innerbib\@empty
\bibitem [{\citenamefont {von Helmholtz}(1895)}]{He95}%
  \BibitemOpen
  \bibfield  {author} {\bibinfo {author} {\bibfnamefont {H.}~\bibnamefont {von
  Helmholtz}},\ }\href@noop {} {\emph {\bibinfo {title} {Wissenschaftliche
  Abhandlungen}}},\ Vol.\ \bibinfo {volume} {Dritter Band}\ (\bibinfo
  {publisher} {Barth, Leipzig},\ \bibinfo {year} {1895})\ \bibinfo {note}
  {http://dx.doi.org/10.3931/e-rara-17433}\BibitemShut {NoStop}%
\bibitem [{\citenamefont {Boltzmann}(1909)}]{Bo84}%
  \BibitemOpen
  \bibfield  {author} {\bibinfo {author} {\bibfnamefont {L.}~\bibnamefont
  {Boltzmann}},\ }\href@noop {} {\emph {\bibinfo {title} {Wissenschaftliche
  Abhandlungen}}},\ Vol.\ \bibinfo {volume} {III. Band}\ (\bibinfo  {publisher}
  {Barth, Leipzig},\ \bibinfo {year} {1909})\ \bibinfo {note}
  {https://phaidra.univie.ac.at/detail/o:6366}\BibitemShut {NoStop}%
\bibitem [{\citenamefont {Brush}(1976)}]{brush1976kind}%
  \BibitemOpen
  \bibfield  {author} {\bibinfo {author} {\bibfnamefont {S.~G.}\ \bibnamefont
  {Brush}},\ }\href@noop {} {\emph {\bibinfo {title} {The kind of motion we
  call heat}}},\ Vol.~\bibinfo {volume} {2}\ (\bibinfo  {publisher}
  {North-Holland Amsterdam},\ \bibinfo {year} {1976})\BibitemShut {NoStop}%
\bibitem [{\citenamefont {Gallavotti}(1999)}]{gallavotti1999statistical}%
  \BibitemOpen
  \bibfield  {author} {\bibinfo {author} {\bibfnamefont {G.}~\bibnamefont
  {Gallavotti}},\ }\href@noop {} {\emph {\bibinfo {title} {Statistical
  mechanics: A short treatise}}}\ (\bibinfo  {publisher} {Springer Science \&
  Business Media},\ \bibinfo {year} {1999})\BibitemShut {NoStop}%
\bibitem [{\citenamefont {Cardin}\ and\ \citenamefont
  {Favretti}(2004)}]{cardin2004helmholtz}%
  \BibitemOpen
  \bibfield  {author} {\bibinfo {author} {\bibfnamefont {F.}~\bibnamefont
  {Cardin}}\ and\ \bibinfo {author} {\bibfnamefont {M.}~\bibnamefont
  {Favretti}},\ }\bibfield  {title} {\bibinfo {title} {On the
  helmholtz-boltzmann thermodynamics of mechanical systems},\ }\href@noop {}
  {\bibfield  {journal} {\bibinfo  {journal} {Continuum mechanics and
  thermodynamics}\ }\textbf {\bibinfo {volume} {16}},\ \bibinfo {pages} {15}
  (\bibinfo {year} {2004})}\BibitemShut {NoStop}%
\bibitem [{\citenamefont {Campisi}(2005)}]{campisi2005mechanical}%
  \BibitemOpen
  \bibfield  {author} {\bibinfo {author} {\bibfnamefont {M.}~\bibnamefont
  {Campisi}},\ }\bibfield  {title} {\bibinfo {title} {On the mechanical
  foundations of thermodynamics: The generalized helmholtz theorem},\
  }\href@noop {} {\bibfield  {journal} {\bibinfo  {journal} {Studies in History
  and Philosophy of Science Part B: Studies in History and Philosophy of Modern
  Physics}\ }\textbf {\bibinfo {volume} {36}},\ \bibinfo {pages} {275}
  (\bibinfo {year} {2005})}\BibitemShut {NoStop}%
\bibitem [{\citenamefont {Campisi}\ and\ \citenamefont
  {Kobe}(2010)}]{campisi2010derivation}%
  \BibitemOpen
  \bibfield  {author} {\bibinfo {author} {\bibfnamefont {M.}~\bibnamefont
  {Campisi}}\ and\ \bibinfo {author} {\bibfnamefont {D.~H.}\ \bibnamefont
  {Kobe}},\ }\bibfield  {title} {\bibinfo {title} {Derivation of the boltzmann
  principle},\ }\href@noop {} {\bibfield  {journal} {\bibinfo  {journal}
  {American Journal of Physics}\ }\textbf {\bibinfo {volume} {78}},\ \bibinfo
  {pages} {608} (\bibinfo {year} {2010})}\BibitemShut {NoStop}%
\bibitem [{\citenamefont {Andrey}(1985)}]{Andrey}%
  \BibitemOpen
  \bibfield  {author} {\bibinfo {author} {\bibfnamefont {L.}~\bibnamefont
  {Andrey}},\ }\bibfield  {title} {\bibinfo {title} {The rate of entropy change
  in non-hamiltonian systems},\ }\href@noop {} {\bibfield  {journal} {\bibinfo
  {journal} {Physics Letters A}\ }\textbf {\bibinfo {volume} {111}},\ \bibinfo
  {pages} {45} (\bibinfo {year} {1985})}\BibitemShut {NoStop}%
\bibitem [{\citenamefont {Evans}\ and\ \citenamefont
  {Rondoni}(2002)}]{RonEvan}%
  \BibitemOpen
  \bibfield  {author} {\bibinfo {author} {\bibfnamefont {D.~J.}\ \bibnamefont
  {Evans}}\ and\ \bibinfo {author} {\bibfnamefont {L.}~\bibnamefont
  {Rondoni}},\ }\bibfield  {title} {\bibinfo {title} {Comments on the entropy
  of nonequilibrium steady states},\ }\href@noop {} {\bibfield  {journal}
  {\bibinfo  {journal} {Journal of Statistical Physics}\ }\textbf {\bibinfo
  {volume} {109}},\ \bibinfo {pages} {895} (\bibinfo {year}
  {2002})}\BibitemShut {NoStop}%
\bibitem [{\citenamefont {Ruelle}(2003)}]{ruelle2003extending}%
  \BibitemOpen
  \bibfield  {author} {\bibinfo {author} {\bibfnamefont {D.~P.}\ \bibnamefont
  {Ruelle}},\ }\bibfield  {title} {\bibinfo {title} {Extending the definition
  of entropy to nonequilibrium steady states},\ }\href@noop {} {\bibfield
  {journal} {\bibinfo  {journal} {Proceedings of the National Academy of
  Sciences}\ }\textbf {\bibinfo {volume} {100}},\ \bibinfo {pages} {3054}
  (\bibinfo {year} {2003})}\BibitemShut {NoStop}%
\bibitem [{\citenamefont {Dunkel}\ and\ \citenamefont
  {Hilbert}(2014)}]{dunkel2014consistent}%
  \BibitemOpen
  \bibfield  {author} {\bibinfo {author} {\bibfnamefont {J.}~\bibnamefont
  {Dunkel}}\ and\ \bibinfo {author} {\bibfnamefont {S.}~\bibnamefont
  {Hilbert}},\ }\bibfield  {title} {\bibinfo {title} {Consistent
  thermostatistics forbids negative absolute temperatures},\ }\href@noop {}
  {\bibfield  {journal} {\bibinfo  {journal} {Nature Physics}\ }\textbf
  {\bibinfo {volume} {10}},\ \bibinfo {pages} {67} (\bibinfo {year}
  {2014})}\BibitemShut {NoStop}%
\bibitem [{\citenamefont {Swendsen}\ and\ \citenamefont
  {Wang}(2016)}]{swendsen2016negative}%
  \BibitemOpen
  \bibfield  {author} {\bibinfo {author} {\bibfnamefont {R.~H.}\ \bibnamefont
  {Swendsen}}\ and\ \bibinfo {author} {\bibfnamefont {J.-S.}\ \bibnamefont
  {Wang}},\ }\bibfield  {title} {\bibinfo {title} {Negative temperatures and
  the definition of entropy},\ }\href@noop {} {\bibfield  {journal} {\bibinfo
  {journal} {Physica A: Statistical Mechanics and its Applications}\ }\textbf
  {\bibinfo {volume} {453}},\ \bibinfo {pages} {24} (\bibinfo {year}
  {2016})}\BibitemShut {NoStop}%
\bibitem [{\citenamefont {Landau}\ and\ \citenamefont
  {Lifshitz}(1976)}]{landau1976mechanics}%
  \BibitemOpen
  \bibfield  {author} {\bibinfo {author} {\bibfnamefont {L.~D.}\ \bibnamefont
  {Landau}}\ and\ \bibinfo {author} {\bibfnamefont {E.~M.}\ \bibnamefont
  {Lifshitz}},\ }\href@noop {} {\emph {\bibinfo {title} {Mechanics: Course of
  theoretical physics, Vol. 1}}},\ Vol.~\bibinfo {volume} {1}\ (\bibinfo
  {publisher} {Butterworth-Heinemann},\ \bibinfo {year} {1976})\BibitemShut
  {NoStop}%
\bibitem [{\citenamefont {Goldstein}\ \emph {et~al.}(2002)\citenamefont
  {Goldstein}, \citenamefont {Poole},\ and\ \citenamefont
  {Safko}}]{goldstein2002classical}%
  \BibitemOpen
  \bibfield  {author} {\bibinfo {author} {\bibfnamefont {H.}~\bibnamefont
  {Goldstein}}, \bibinfo {author} {\bibfnamefont {C.}~\bibnamefont {Poole}},\
  and\ \bibinfo {author} {\bibfnamefont {J.}~\bibnamefont {Safko}},\
  }\href@noop {} {\bibinfo {title} {Classical mechanics}} (\bibinfo {year}
  {2002})\BibitemShut {NoStop}%
\bibitem [{\citenamefont {Callen}(1998)}]{callen1998thermodynamics}%
  \BibitemOpen
  \bibfield  {author} {\bibinfo {author} {\bibfnamefont {H.~B.}\ \bibnamefont
  {Callen}},\ }\href@noop {} {\bibinfo {title} {Thermodynamics and an
  introduction to thermostatistics}} (\bibinfo {year} {1998})\BibitemShut
  {NoStop}%
\bibitem [{\citenamefont {Solon}\ \emph {et~al.}(2015)\citenamefont {Solon},
  \citenamefont {Fily}, \citenamefont {Baskaran}, \citenamefont {Cates},
  \citenamefont {Kafri}, \citenamefont {Kardar},\ and\ \citenamefont
  {Tailleur}}]{solon2015pressure}%
  \BibitemOpen
  \bibfield  {author} {\bibinfo {author} {\bibfnamefont {A.~P.}\ \bibnamefont
  {Solon}}, \bibinfo {author} {\bibfnamefont {Y.}~\bibnamefont {Fily}},
  \bibinfo {author} {\bibfnamefont {A.}~\bibnamefont {Baskaran}}, \bibinfo
  {author} {\bibfnamefont {M.~E.}\ \bibnamefont {Cates}}, \bibinfo {author}
  {\bibfnamefont {Y.}~\bibnamefont {Kafri}}, \bibinfo {author} {\bibfnamefont
  {M.}~\bibnamefont {Kardar}},\ and\ \bibinfo {author} {\bibfnamefont
  {J.}~\bibnamefont {Tailleur}},\ }\bibfield  {title} {\bibinfo {title}
  {Pressure is not a state function for generic active fluids},\ }\href@noop {}
  {\bibfield  {journal} {\bibinfo  {journal} {Nature Physics}\ }\textbf
  {\bibinfo {volume} {11}},\ \bibinfo {pages} {673} (\bibinfo {year}
  {2015})}\BibitemShut {NoStop}%
\bibitem [{\citenamefont {Landau}\ and\ \citenamefont
  {Lifshitz}(2013)}]{landau2013course}%
  \BibitemOpen
  \bibfield  {author} {\bibinfo {author} {\bibfnamefont {L.~D.}\ \bibnamefont
  {Landau}}\ and\ \bibinfo {author} {\bibfnamefont {E.~M.}\ \bibnamefont
  {Lifshitz}},\ }\href@noop {} {\emph {\bibinfo {title} {Statistical Physics.
  Course of theoretical physics, Vol 5}}}\ (\bibinfo  {publisher} {Elsevier},\
  \bibinfo {year} {2013})\BibitemShut {NoStop}%
\bibitem [{\citenamefont {Rondoni}\ and\ \citenamefont
  {Cohen}(2002)}]{rondoni2002some}%
  \BibitemOpen
  \bibfield  {author} {\bibinfo {author} {\bibfnamefont {L.}~\bibnamefont
  {Rondoni}}\ and\ \bibinfo {author} {\bibfnamefont {E.}~\bibnamefont
  {Cohen}},\ }\bibfield  {title} {\bibinfo {title} {On some derivations of
  irreversible thermodynamics from dynamical systems theory},\ }\href@noop {}
  {\bibfield  {journal} {\bibinfo  {journal} {Physica D: Nonlinear Phenomena}\
  }\textbf {\bibinfo {volume} {168}},\ \bibinfo {pages} {341} (\bibinfo {year}
  {2002})}\BibitemShut {NoStop}%
\bibitem [{\citenamefont {Rondoni}(2021)}]{rondoni2021introduction}%
  \BibitemOpen
  \bibfield  {author} {\bibinfo {author} {\bibfnamefont {L.}~\bibnamefont
  {Rondoni}},\ }\bibfield  {title} {\bibinfo {title} {Introduction to
  nonequilibrium statistical physics and its foundations},\ }\href@noop {}
  {\bibfield  {journal} {\bibinfo  {journal} {Frontiers and Progress of Current
  Soft Matter Research}\ ,\ \bibinfo {pages} {1}} (\bibinfo {year}
  {2021})}\BibitemShut {NoStop}%
\bibitem [{\citenamefont {Arnol'd}(2013)}]{arnol2013mathematical}%
  \BibitemOpen
  \bibfield  {author} {\bibinfo {author} {\bibfnamefont {V.~I.}\ \bibnamefont
  {Arnol'd}},\ }\href@noop {} {\emph {\bibinfo {title} {Mathematical methods of
  classical mechanics}}},\ Vol.~\bibinfo {volume} {60}\ (\bibinfo  {publisher}
  {Springer Science \& Business Media},\ \bibinfo {year} {2013})\BibitemShut
  {NoStop}%
\bibitem [{\citenamefont {Hoover}(1993)}]{hoover1993nonequilibrium}%
  \BibitemOpen
  \bibfield  {author} {\bibinfo {author} {\bibfnamefont {W.~G.}\ \bibnamefont
  {Hoover}},\ }\bibfield  {title} {\bibinfo {title} {Nonequilibrium molecular
  dynamics: the first 25 years},\ }\href@noop {} {\bibfield  {journal}
  {\bibinfo  {journal} {Physica A: Statistical Mechanics and its Applications}\
  }\textbf {\bibinfo {volume} {194}},\ \bibinfo {pages} {450} (\bibinfo {year}
  {1993})}\BibitemShut {NoStop}%
\bibitem [{\citenamefont {Hoover}(2012)}]{hoover2012computational}%
  \BibitemOpen
  \bibfield  {author} {\bibinfo {author} {\bibfnamefont {W.~G.}\ \bibnamefont
  {Hoover}},\ }\href@noop {} {\emph {\bibinfo {title} {Computational
  statistical mechanics}}}\ (\bibinfo  {publisher} {Elsevier},\ \bibinfo {year}
  {2012})\BibitemShut {NoStop}%
\bibitem [{\citenamefont {Rapaport}\ and\ \citenamefont
  {Rapaport}(2004)}]{rapaport2004art}%
  \BibitemOpen
  \bibfield  {author} {\bibinfo {author} {\bibfnamefont {D.~C.}\ \bibnamefont
  {Rapaport}}\ and\ \bibinfo {author} {\bibfnamefont {D.~C.~R.}\ \bibnamefont
  {Rapaport}},\ }\href@noop {} {\emph {\bibinfo {title} {The art of molecular
  dynamics simulation}}}\ (\bibinfo  {publisher} {Cambridge university press},\
  \bibinfo {year} {2004})\BibitemShut {NoStop}%
\bibitem [{\citenamefont {J~Evans}\ and\ \citenamefont
  {P~Morriss}(2008)}]{j2007statistical}%
  \BibitemOpen
  \bibfield  {author} {\bibinfo {author} {\bibfnamefont {D.}~\bibnamefont
  {J~Evans}}\ and\ \bibinfo {author} {\bibfnamefont {G.}~\bibnamefont
  {P~Morriss}},\ }\href@noop {} {\emph {\bibinfo {title} {Statistical mechanics
  of nonequilbrium liquids}}}\ (\bibinfo  {publisher} {Cambridge university
  press},\ \bibinfo {year} {2008})\BibitemShut {NoStop}%
\bibitem [{\citenamefont {Jepps}\ and\ \citenamefont
  {Rondoni}(2010)}]{jepps2010deterministic}%
  \BibitemOpen
  \bibfield  {author} {\bibinfo {author} {\bibfnamefont {O.~G.}\ \bibnamefont
  {Jepps}}\ and\ \bibinfo {author} {\bibfnamefont {L.}~\bibnamefont
  {Rondoni}},\ }\bibfield  {title} {\bibinfo {title} {Deterministic
  thermostats, theories of nonequilibrium systems and parallels with the
  ergodic condition},\ }\href@noop {} {\bibfield  {journal} {\bibinfo
  {journal} {Journal of Physics A: Mathematical and Theoretical}\ }\textbf
  {\bibinfo {volume} {43}},\ \bibinfo {pages} {133001} (\bibinfo {year}
  {2010})}\BibitemShut {NoStop}%
\bibitem [{\citenamefont {Todd}\ and\ \citenamefont
  {Daivis}(2017)}]{todd2017nonequilibrium}%
  \BibitemOpen
  \bibfield  {author} {\bibinfo {author} {\bibfnamefont {B.~D.}\ \bibnamefont
  {Todd}}\ and\ \bibinfo {author} {\bibfnamefont {P.~J.}\ \bibnamefont
  {Daivis}},\ }\href@noop {} {\emph {\bibinfo {title} {Nonequilibrium molecular
  dynamics: theory, algorithms and applications}}}\ (\bibinfo  {publisher}
  {Cambridge University Press},\ \bibinfo {year} {2017})\BibitemShut {NoStop}%
\bibitem [{\citenamefont {Kirkwood}(1935)}]{kirkwood1935statistical}%
  \BibitemOpen
  \bibfield  {author} {\bibinfo {author} {\bibfnamefont {J.~G.}\ \bibnamefont
  {Kirkwood}},\ }\bibfield  {title} {\bibinfo {title} {Statistical mechanics of
  fluid mixtures},\ }\href@noop {} {\bibfield  {journal} {\bibinfo  {journal}
  {The Journal of chemical physics}\ }\textbf {\bibinfo {volume} {3}},\
  \bibinfo {pages} {300} (\bibinfo {year} {1935})}\BibitemShut {NoStop}%
\bibitem [{\citenamefont {Jarzynski}(2017)}]{jarzynski2017stochastic}%
  \BibitemOpen
  \bibfield  {author} {\bibinfo {author} {\bibfnamefont {C.}~\bibnamefont
  {Jarzynski}},\ }\bibfield  {title} {\bibinfo {title} {Stochastic and
  macroscopic thermodynamics of strongly coupled systems},\ }\href@noop {}
  {\bibfield  {journal} {\bibinfo  {journal} {Physical Review X}\ }\textbf
  {\bibinfo {volume} {7}},\ \bibinfo {pages} {011008} (\bibinfo {year}
  {2017})}\BibitemShut {NoStop}%
\end{thebibliography}%
\end{document}